%% file: icdcs_main.tex
\documentclass[conference]{IEEEtran}
\IEEEoverridecommandlockouts
\usepackage{amsmath,amssymb,amsfonts}
\usepackage{algorithmic}
\usepackage{graphicx}
\usepackage{textcomp}
\usepackage{xcolor}

\usepackage{algorithm}
\usepackage{mathtools}
\usepackage{tabularx}
\usepackage{multirow}
\usepackage{subfigure}
\usepackage{color}

\def\BibTeX{{\rm B\kern-.05em{\sc i\kern-.025em b}\kern-.08em
    T\kern-.1667em\lower.7ex\hbox{E}\kern-.125emX}}
\begin{document}

\title{Practical Privacy Preservation in a Mobile Cloud Environment}

\author{
\IEEEauthorblockN{Dimitrios Tomaras, Michail Tsenos, Vana Kalogeraki}
\IEEEauthorblockA{Department of Informatics\\
Athens University of Economics and Business, Athens, Greece\\
\{tomaras, tsemike, vana\}@aueb.gr}
}

\maketitle

\begin{abstract}
The proliferation of smartphone devices has led to the emergence of powerful user services from enabling interactions with friends and business associates to mapping, finding nearby businesses and alerting users in real-time. Moreover, users do not realize that continuously sharing their trajectory data with online systems may end up revealing a great amount of information in terms of their behavior, mobility patterns and social relationships. Thus, addressing these privacy risks is a fundamental challenge. In this work, we present $TP^3$, a Privacy Protection system for Trajectory analytics. Our contributions are the following: (1) we model a new type of attack, namely 'social link exploitation attack', (2) we utilize the coresets theory, a fast and accurate technique which approximates well the original data using a small data set, and running queries on the coreset produces similar results to the original data, and (3) we employ the Serverless computing paradigm to accommodate a set of privacy operations for achieving high system performance with minimized provisioning costs, while preserving the users' privacy. We have developed these techniques in our $TP^3$ system that works with state-of-the-art trajectory analytics apps and applies different types of privacy operations. Our detailed experimental evaluation illustrates that our approach is both efficient and practical.
\end{abstract}


\input{introduction}

\input{model}

\input{approach}

\input{experiments}

\input{related}

\input{conclusion}

\section*{Acknowledgment}
This research has been supported by the H2020 LAMBDA Project 734242, the EU ICT-48 2020 project TAILOR (No. 952215) and the H2020 AutoFair project (No. 101070568).

\bibliographystyle{IEEEtran}

\bibliography{biblio}
% \end{thebibliography}
% \vspace{12pt}
% \color{red}
% IEEE conference templates contain guidance text for composing and formatting conference papers. Please ensure that all template text is removed from your conference paper prior to submission to the conference. Failure to remove the template text from your paper may result in your paper not being published.

\end{document}

%% file: introduction.tex
\section{Introduction}

The proliferation of smartphone devices has opened up a new era of collaboration and sharing. With the advent of the Internet of Things (IoT), the paradigm shift towards interconnected devices and platforms has allowed the gathering of more information about the environment from heterogeneous sources and the exchange of information with the real world. The IoT paradigm allows for a renewed form of context-aware computing, 
%Vana_1: den katalabainw ti ennoeis "set up the pedestal for a renewed form of .."? 
%Dimitris: Addressed. To allaksa se principles. Ennooume tis basikes arxes.
where applications interact with the user, and adapt their services based on the prevailing user context.
%\footnote{https://www.uber.com},\footnote{https://www.swarmapp.com/},\footnote{https://triplogmileage.com/}, 
For instance, trajectory-based systems such as Uber (https://www.uber.com), Foursquare's Swarm (https://www.swarmapp.com/) and DiDi (https://www.didiglobal.com/) 
%Vana3: bale references/footnotes se ola ta systhmata, kai oxi mono sto Runkeeper
have enabled users to systematically share updates about their activities and whereabouts. %during outdoor promenades. 
In these systems, users often contribute their user IDs and timestamped locations in real-time, possibly enriched with multimedia content, such as images, videos or text, in order to track family members and friends, get rewards, or receive recommendations about places of interest. 
For the vast majority of the individuals, there are several benefits from sharing their whereabouts 
%during \textit{promenades}
(e.g., friends trying to find each other in busy places such as shopping centers or parks, parents tracking their children locations for safety purposes or citizens earning free parking time in smart cities\footnote{http://www.vavel-project.eu/}).

% Obviously, such systems play a significant role through the connectivity they provide and the information they deliver; this is typically exploited to provide their users with ever-improving services and better user experience. In particular, numerous applications employ data analysts to exploit such systems for delivering even better services, tailored to the needs of users, such as better profiles for dating apps or additional discounts for apps that connect users with local merchants. 

{\bf Trajectory Analytics.}
%The growing ubiquity of such systems has blurred the distinction between online and offline social networks. 
%The distributed nature of the data sources (i.e. different interconnected apps installed on users' mobile phones), the volume of the data produced by these apps and the richness of information they encapsulate, are the key players that call for more advanced paradigms for data processing. 
% Current research has focused on the development of system frameworks that enable studying a wide variety of human activities. 
It has been shown that users issue various types of \emph{queries} to trajectory analytics applications while taking private leisure walks in public places or while strolling with friends around the city, asking for recommendations or other types of services, e.g. querying for available taxis, which are performed by sharing their own trajectories.
The majority of these state-of-the-art works aim at capturing human movement, dynamics, regularities \cite{backstrom2010find} and provide secure analytics \cite{beck2017privapprox} observing mobility patterns in data streams, providing answers to questions regarding the geographic movement of people ({\it "where do they move?"}) or aiming to understand how social ties impact mobility patterns, answering questions such as {\it "what the friends of a user can reveal about the user's mobility patterns"} and {\it "how similar users are based on their mobility"} \cite{pham2013ebm,backes2017walk2friends}. 
A fundamental insight in these works is that people exhibit strong periodic behavior in their mobility patterns 
as they move back and forth between their homes and workplaces \cite{Cho2011,Boutsis2016a}, and
that user mobility is shaped by our social relationships as we are more likely to visit places that our friends 
and people with similar interests have visited in the past or have shared common activities with,
a phenomenon known as social homophily \cite{tang2014mining}. 
%DIMITRIS_MDM_22: Tin epomeni protasi tin evala se comment giati mou fainetai oti einai disjoint me ayto pou leme parapanw. Den tonizoume i perigrafoume kapoio point gia real-time low latency. To point tis paragraph einai na poume oti me ta trajectory analytics meletame to pws kineitai o xristis kai pws diamorfwnetai to mobility me tous filous tou.
% However, these works typically deal with single-shot batch queries and cannot be used for real-time low-latency trajectory analytics.
%
%the user mobility patterns, as depicted in the usage of these systems, are often shaped and influenced by the users' social connections. 
%Typically, in their daily routine, users of such systems take "\textit{promenades}" either in the form of private leisure walks in public places so as to meet others, or in the form of strolling with friends around the city's areas or even following up their friends' activities during their own promenades. 
%In particular, users are more likely to promenade city streets choosing to pass through areas where they have shared common activities with their friends.
% However, friends exhibit stronger similarities in their mobility profiles compared to strangers. %Social relationships can disclose highly privacy-sensitive information that is deeply connected with the individuals' social identities\cite{backes2017walk2friends}
%or reveal sensitive user locations (e.g. home, work).  

{\bf Trajectory Privacy Preservation.}
Current trajectory analytics applications are crude, since users are asked to "opt-in" (thus allowing for disclosing possible sensitive data to them) in order to receive higher accuracy, whereas in others they may "opt-out" and accept generic recommendations but with much lower quality and accuracy.
A typical operating scenario of such applications employs a system for delivering trajectories from users to data analysts, in order to be further processed and returns a set of beneficial services for the users. In the current big data ecosystem, it is typical to have direct access to users’ private data, and they must be trusted not to abuse it. However, this trust has been violated in the past\footnote{https://www.eff.org/deeplinks/2015/01/healthcare.gov-sends-personal-data}.
Additionally, as privacy preferences are subjective by nature, only a small percentage of the users of these systems realize the serious privacy implications that may arise and their extent. 
%These systems obviously raise serious privacy concerns.
An adversary, or third party, can extract social trajectory-based data to identify social ties among the users of these systems. 
The adversary can further exploit the extracted information to target \emph{individual users} for marketing campaigns, 
monitor their movements to compromise one's personal safety, or even act on behalf of a third company ({\it i.e.}, an insurance company that aims to extract personal biometric characteristics from fitness applications, evaluate the
health status of the users and appropriately adjust their insurance rates \cite{fereidooni2017fitness}).
The adversary can also target \emph{groups of users with similar mobility patterns}.
For example, it is known that location and mobility data are examined by NSA to identify new (or unknown) members of criminal gangs or terrorist cells 
of targets that already knows about\footnote{https://www.expressvpn.com/internet-privacy/guides/nsa-spying/}. 
Or finally, groups of users can be exploited for targeted advertising: 
%The user's mobility pattern similarity to a given user group could help increase the accuracy in advertisement recommendation. F
advertising companies create targeted ad groups, based on interests, lifestyles, demographics, geo-location or mobility patterns.
%so that consumers who are likely to have a strong preference will receive the message instead of those who have no interest and whose preferences do not match a product's attribute.
%for example, if a user's mobility pattern matches the mobility pattern of a teenager user group, then the system can pop up advertisements of interest to that particular age group or recommend geo-located activities or interest to that group.
Similar examples can be found in mobile alerting systems or social transportation systems \cite{zheng2016big} where mobility groups are utilized for ride-sharing or to react to disruptions of transportation services in real-time.

{\bf The serverless computing model.} 
%In this work, we make the observation that the two computing paradigms, i.e. trajectory privacy preservation and low-latency trajectory analytics, are complementary. Both computing paradigms strive for enhancing the user experience and privacy. 
Recently, a new privacy-preserving model for trajectories in Mobile Cloud Environments(MCEs) has been proposed,
referred to as on-device public-private TPP model \cite{epasto2019device}, 
where user data is partitioned into two parts: (a) a public part that can be shared with a data analyst, and (b) a private part (on-device data) that is protected from disclosure.
The model conducts computations using both private and public data on the user's phone. However, storing all the user trajectories in the phones is impossible due to storage limitations and therefore, unavoidably, they must be conducted through a \emph{mobile cloud service provider} before being disclosed to the data analyst. The advent of cloud computing has enabled providing low-latency trajectory analytics and has enabled many service providers to move their hardware infrastructure from on-premise deployments into large-scale shared cloud resources \cite{lin2020serverless}. The \emph{Serverless Computing} has been introduced recently in the literature \cite{lin2020serverless} as a scheme to enable Mobile Cloud providers balance the \emph{trade-off} between the computational costs for users (i.e. provide results with low-latency) and the provisioning costs (i.e resource allocation costs), compared to traditional on-premises deployments. While marrying these two paradigms, we face two challenges: the first is how to detect social ties between users based on the trajectories, in real-time. The question is whether the estimation of privacy exposure can utilize only a subset (sample) of the user's trajectories. The second challenge is, since the Serverless Computing paradigm enables for a \textit{pay-as-you-use} service, how we can exploit it to balance the trade-off between providing trajectory analytics with low-latency and minimizing the costs associated with using such a service, while considering the different privacy perspectives of the individual users.

The goal of the work is to present a cost-effective and practical system for MCEs enhanced with a user-tunable degree of privacy preservation, that provide solutions to these two major challenges: (1) minimize the volume of mobility patterns required for the privacy estimation analysis to enable on-device storage, and (2) utilize the serverless computing paradigm for trajectory analytics,
to balance the trade-off between low-latency computations and minimization of provisioning costs, preserving at the same time the privacy of the trajectories.

{\bf Contributions.} In this work we make the following contributions:
\begin{itemize}
    \item We propose {\bf $TP^{3}$}, a Practical Privacy preservation system for trajectory analytics in MCEs.
    $TP^{3}$ adopts an on-device model that \textit{reduces the volume of the examined mobility patterns} using the theory of coresets and allows the system to obtain guaranteed, fast and accurate approximations of user trajectory data. 

    \item We model a new type of attack, namely \textit{social link exploitation attack}, where a third party data analyst can infer information about the user and the user's behavior by associating the user's patterns with groups of similar users while sharing trajectory data. 

    \item We exploit the Serverless Computing paradigm for providing a balance between high performance trajectory analytics
    %Vana3: na bgaleis to high-availability se ayto to contribution? 
    %Dimitris: addressed
    and low provisioning costs, utilizing a Pareto-Frontier search algorithm. %, while achieving trajectory privacy. 
    We  apply four different privacy operations achieving a good balance between accuracy and privacy of the disclosed trajectories. 
    
    \item We have implemented the $TP^{3}$ system to support trajectory analytics apps on top of Android devices and the OpenFaaS serverless platform, and evaluated its performance. %utilizing four different types of privacy operations.
    We illustrate %its privacy preservation efficiency, which 
    that $TP^{3}$ is a cost-effective approach and can achieve at least 47\% reduction of the risk of privacy exposure while users are able to contribute data to trajectory analytics apps.

\end{itemize}

%The rest of the paper is structured as follows: In section 2, we provide some preliminaries including our System Model and we define our problem. In section 3, we extensively describe our approach. In section 4, we present our experimental evaluation. Section 5 describes related work and finally, in section 6, we conclude with lessons learnt from this work.

%% file: model.tex
\section{Model and Threat Definition}

%In this section we first present our system model and our assumptions, discuss the threat model and finally describe our problem.

\subsection{On-device Model}

{\bf Users. } Users $\mathcal{U}$ in the $TP^{3}$ system are characterized by the tuple $\langle id_{u}, \{p_{u,i}^{\tau}\}\rangle$, where $id_{u}$ is the unique id of user $u$ and $\{p_{u,i}^{\tau}\}$ is a list of all the contributed publicly available data reports made by the user through the trajectory analytics apps. Typically, these reports are coupled with location and timestamps and 
%are made available to other users to 
illustrate the user's presence activity. 
%These may vary from presence reports ({\it i.e.,} in trajectory-based social network applications) to fitness summary reports ({\it i.e.,} in fitness applications).
The $i$-th spatio-temporal data report issued by a user $u$, $p_{u,i}^{\tau}$, is a tuple  $\langle lat_{i}, lon_{i} , \tau, dat_{u,i,\tau} \rangle$, where $lat_{i}, lon_{i}$ are the \textit{geospatial coordinates}, $\tau$ a \textit{timestamp} value annotating the time when the user $u$ issued the specific report and $dat_{u,i,\tau}$ denotes the data associated with the spatio-temporal report. In the case of a traffic monitoring application e.g., the data can represent \textit{the traffic state} (e.g. "traffic jam", "road closure" etc.), whereas in location-based recommendation apps it can reveal \textit{the location semantics} of a place where a user has checked-in.

{\bf Trajectories. } Users share reports to trajectory analytics apps which are represented in the form of trajectories. That is, users maintain locally their spatio-temporal data reports which form their mobility patterns. Each user $u$ trajectory $tr_{u}^{l}$ has a unique identity $l$ and a sequence of spatio-temporal data reports $p_{u,i}^{\tau}$; these denote the route the user was following while issuing the spatio-temporal data reports.

{\bf Mobility Profiles.} The Mobility Profile (MP) of a user $u$, 
%Vananew: giati exeis user u_1 kai oxi apla user u?
%DimitrisQ Addressed. to eixame apo ta palia chains twra einai idio me to ypoloipo paper.
$\mathcal{G}_{u}$, is represented through the user's trajectories,
%. More formally,
$  \mathcal{G}_{u}=\langle \{tr_{u}^{l_{1}},tr_{u}^{l_{2}},...,tr_{u}^{l_{\kappa}}\}\rangle$. 
We discuss in detail how mobility profiles are compiled in our $TP^3$ system in section \ref{sec:mobiprof}.
%Vananew: is the above section number correct?
%Dimitris: Addressed. Evala label kai sto antistoixo section sto approach.

{\bf Utility.} The notion of utility has been introduced in recent works\cite{gursoy2018utility,xiao2015protecting} as a performance metric of location privacy protection systems in order to capture the trade-off between data quality and user privacy preservation.
The goal is to preserve data utility as much as possible 
(in terms of how useful and accurate the data report is) 
%Vana5: Dimitri, des thn parapanw parenthesi, einai ok?
%Dimitris: to usefulness me mperdeuei,to allaksa se how much useful and accurate the data report is. Den mou kathotan kala ypo tin ennoia oti den orizotan to usefulness.
%while protecting the users against the social link exploitation attack.
%Vananew: allaxa to parapanw me thn akolouthi protash. den exoun eisagei akoma to attack
%Dimitris: Addressed. To eida, kai ontws den exoume eisagei to attack.
while preserving the users' privacy.
In $TP^{3}$, our aim is to measure the utility $UT$ of a perturbed trajectory $\hat{tr}_{u}^{l}$ , thus we define it as the inverted distance between each one true report $p_{u,i}^{\tau}\in tr_{u}^{l}$ and the corresponding released report $\hat{p}_{u,i}^{\tau}\in \hat{tr}_{u}^{l}$. More formally,
$
    UT(\hat{tr}_{u}^{l},tr_{u}^{l}) = \sum_{\substack{\forall p_{u,i}^{\tau}\in tr_{u}^{l} \\ \forall \hat{p}_{u,i}^{\hat{\tau}}\in \hat{tr}_{u}^{l}}}\frac{1}{\sqrt{E \parallel\hat{p}_{u,i}^{\hat{\tau}}-p_{u,i}^{\tau}\parallel_{2}^{2}}}
$,
where $E\parallel \bullet \parallel$ denotes the difference between the reports (also called "correctness" \cite{shokri2011quantifying}). In the case of \textit{trajectory analytics applications} the aforementioned formula considers \textit{the geographical distance} between the reports, whereas in the case of \textit{emergency response trajectory analytics applications}, for which time is a critical parameter, the equation also considers \textit{the difference in the timestamps} among the original and the released reports.

{\bf Social Graph. } The set of users $\mathcal{U}$ form an undirected social graph $G=(V,E,\mathcal{S})$, where each node $v_{u}\in V$ denotes user $u$, each edge $e_{mn}\in E$ annotates the social tie between users $u_{m}$ and $u_{n}$, and finally $\mathcal{S}(u_{m},u_{n})\in \mathcal{S}$ is a value that represents the strength of the social tie between users $u_{m}$ and $u_{n}$.

%{\bf On-device public/private data. } The on-device paradigm sets the requirement of partioning the user data in a private and a public part. 

%DIM_ICDCS_2021: To evala se comment giati mou fainotan unrelated me to scope tou systimatos.
% In $TP^{3}$, we adopt an \emph{on-device public/private data model} as a one of the building blocks of our LPPS, where we consider as private data the social graph associated with the user. The user keeps his social connections, that is, his own social graph that he is not willing to share, locally as private data. 
% %Vananew: to paapanw den to katalabainw, pws o user identifies his own social graph? kai pws kanei identify that these are private data sto app? den to katalabainw pws phgainetai pragmatika se mia efarmogh
% %Dimitris: To pio aplo paradeigma se auto tha mporouse na einai to contact list to opoio tha prepei na kanei share o xristis gia na brei poioi alloi filoi tou einai mesa stin efarmogi.
% As the user shares data reports (which are considered as public data shared by the user) using $TP^{3}$, an appropriate on-device service collects the reports and forwards them to a central service for further evaluation. These data reports are the real reports which will be evaluated for being distorted or not.

\subsection{Serverless Model}
%mixali, svinw to serverless privacy functions edw gia an min se mperdeuei. Dinoume kateutheian ton orismo tou serverless model opws eipame sto skype kai tha to grapsw katallila sto approach.
We chose to deploy $TP^3$ on a Serverless environment rather than a traditional hosting solution in a cloud provider. %The SPFs are a set of distributed privacy functions for ingesting the trajectories, shared by the users, which represent the user's mobility patterns, evaluating the user's similarity against other users and sanitize them before disclosure. Upon completing those steps, the SPF will forward the trajectories to the respective data analyst that wishes to exploit them. 
Choosing the Serverless computing model \cite{lin2020serverless}, we have a lower operational and deployment cost due to its unique pricing policy based on a \emph{pay-as-you-use} model. 
% In traditional host solutions, it is typical to rent some dedicated virtual or physical machines for a specified amount of time. In the Serverless model we can use 
This model allows for using ephemeral containers that are utilized only during our workload. During long periods of inactivity, containers stop running automatically (scale to zero) to keep the operational cost low, since it provides a seamless method for autoscaling the resources. 
%There is no need to have dedicated system administrators or custom built automated tools to manage the existing infrastructure. 
%without the need of knowing the internals of the underlying technologies. 
That is the number of active instances can adapt dynamically according to the number of requests, e.g., if 3 active instances are typically employed, these can increase to 10 during data bursts and can go back to normal levels when
the number of requests decreases. 
Apart from the aforementioned benefits, it allows developers to build simpler software by designing their services as functions. In our approach we use OpenFaaS (https://www.openfaas.com/) as our Serverless environment, but our system works inline with existing state-of-the art Serverless systems such as AWS, GCF, IBM Cloud, etc.

\begin{figure*}[t!]\centering
\begin{minipage}{0.725\linewidth}\centering
\includegraphics[width=\linewidth]{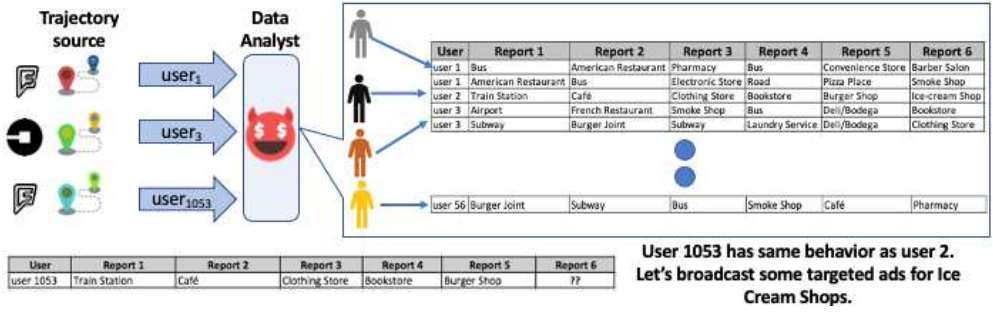}
\caption{Social Strength Exploitation Attack Example.}\label{fig:example}
\end{minipage}\hfill
\end{figure*}

\subsection{Threat Model}

{\bf Social Strength. }  To infer whether a pair of individuals are socially connected, recent works \cite{pham2013ebm,backes2017walk2friends} have attempted to model the relationship between the users' mobility patterns and their social ties. 
One approach is to use a graph model \cite{backes2017walk2friends} where learning the users' mobility features and evaluating the users' mobility similarity, the social tie between two users can be captured.
A similar model was introduced in \cite{pham2013ebm} where they aim at recording the diversity of the locations visited by the users. However, both approaches are not appropriate for our setting as they focus on single locations rather than trajectories, and thus cannot encapsulate the users' movement regularity, dependencies or transitions among different locations. 
Trajectory-based modeling, on the other hand, allows us to meet with high accuracy the requirements of applications 
with spatiotemporal correlations among queries issued by the users for trajectory analytics apps \cite{zang2011anonymization}.

We use an entropy-based model to capture the social strength among two users of the system, given their mobility profiles. Our metric captures the diversity of the users (i.e. in terms of the number of different places they visit) by evaluating the similarity of their mobility profiles. 
Thus, the social strength $\mathcal{S}(\mathcal{G}_{u_{1}},\mathcal{G}_{u_{2}})$ among two users $u_{1},u_{2}$ given their mobility profiles $\mathcal{G}_{u_{1}}$, $\mathcal{G}_{u_{2}}$, is defined 
as:
%as a linear function that describes the diversity of the two users. Specifically,
$
  \mathcal{S}(\mathcal{G}_{u_{1}},\mathcal{G}_{u_{2}})=\alpha \cdot e^{\mathcal{H}(\mathcal{G}_{u_{1}}|\mathcal{G}_{u_{2}})}
$
where
% \begin{align}\label{eq:entropy}
$  \mathcal{H}(\mathcal{G}_{u_{1}}|\mathcal{G}_{u_{2}})=-\sum_{\substack{1 \leq z \leq |\mathcal{G}_{u_{1}}|\\ 1 \leq y \leq |\mathcal{G}_{u_{2}}|}}\mathcal{P}(tr_{u_{1}}^{l_{z}}|tr_{u_{2}}^{l_{y}})\cdot log \mathcal{P}(tr_{u_{1}}^{l_{z}}|tr_{u_{2}}^{l_{y}})
$
% \end{align}
% \vspace{-0.5em}
% \begin{figure}[t!] \centering
% \begin{minipage}{0.8\linewidth}
% \centering
% \includegraphics[width=0.9\textwidth]{percomScenario.png}
% \caption{Scenario.}\label{fig:scenario}
% \end{minipage}
% \end{figure}

\noindent
where $\alpha \in (0,1]$ is a parameter to tune the degree of privacy preservation tailored to the user's personal needs (explained in detail in the next section). For instance, a value of ($\alpha = 1$) indicates that the user is highly concerned about the privacy. Additionally, $\mathcal{S}(\mathcal{G}_{u_{1}},\mathcal{G}_{u_{2}})\in (0,1]$; a value of $\mathcal{S}(\mathcal{G}_{u_{1}},\mathcal{G}_{u_{2}})$ equal to 1 indicates that the two users are highly socially connected and have identical mobility behavior, while a value close to 0 indicates that there is little social connection among the users.

The probability function $\mathcal{P}(tr_{u_{1}}^{l_{z}}|tr_{u_{2}}^{l_{y}})$ evaluates the similarity between the trajectories $tr_{u_{1}}^{l_{z}}$, $tr_{u_{2}}^{l_{y}}$ of users $u_{1}$,$u_{2}$. 
To perform this computation, we utilize the notion of \textit{expected edit distance metric \cite{cotterell2014stochastic}, which characterizes the similarity between trajectories with a respective probability to show up.} That is, given the trajectory $tr_{u_{2}}^{l_{y}}$, the function captures what is the probability to find the trajectory $tr_{u_{1}}^{l_{z}}$, as implied by the metric. Therefore, the probability function $\mathcal{P}(tr_{u_{1}}^{l_{z}}|tr_{u_{2}}^{l_{y}})$ is defined formally as:
% \begin{align}\label{eq:prob}
$    \mathcal{P}(tr_{u_{1}}^{l_{z}}|tr_{u_{2}}^{l_{y}})=\frac{LCS(tr_{u_{1}}^{l_{z}}, tr_{u_{2}}^{l_{y}})}{|tr_{u_{1}}^{l_{z}}|}
$
% \end{align}
where $LCS(\bullet,\bullet)$ denotes the longest common consecutive subsequence between two trajectories. We assume a third party data analyst that aims at exploiting the social ties among mobile users. The data analyst has compiled a list $\mathcal{D}$ of mobility profiles (MPs) for all users that have shared trajectories with the trajectory analytics apps. 
%during their promenades. 
The MPs are compiled from the users' mobility patterns using most frequent subset pattern mining procedure\cite{fournier2013tks}. 

{\bf Social Link Exploitation Attack}.
We introduce a novel type of attack, namely the \emph{social link exploitation attack}, where the a data analyst attempts to exploit the social relationships through trajectory analytics apps in order to predict the movement patterns of a particular user. In our setting, the analyst has already compiled 
a model of human mobility based on the geospatial movements of the users. % and their social connections.
By estimating the social strength between a user and the set of compiled mobility profiles, the attacker is able to associate the user with other users with the same behavior and estimate the user's mobility pattern \cite{kong2015privacy}.
The attacker can exploit this information not only to compromise one's safety but also for marketing purposes to predict the most likely venue that a user will visit and 
select an appropriate ad to pop-up in the user's phone.

{\bf Example.} 
%We provide a real-world scenario which depicts the importance of building an appropriate system for preserving users' privacy against the curious and possibly malicious behavior of a data analyst. 
In Figure \ref{fig:example} we illustrate a scenario of how a curious data analyst can associate a user $u$ with one of the MPs of the set $\mathcal{D}$ that has already been compiled. %of users $\mathcal{B}$.
The left part of the figure illustrates three users of a specific trajectory analytics app. In a real world scenario, the users typically provide direct access of their trajectories to the data analyst.
% log in to that app using the account credentials of their location-based apps. By doing so, they subsequently provide access to their spatio-temporal reports from that app, 
% %Vana_1: ti ennoeis "This has subsequently given access to their spatio-temporal reports..? Poios einai to "this" ? 
% %Dimitris: Addressed.
% since, any time they use it, it is able to capture the locations visited. 
The right part of the figure shows a small sample of the MPs compiled from our Fousquare dataset (described in section \ref{sec:evaluation}) for various users utilizing this app.
We show 6 different trajectories that correspond to three users, based on the Foursquare data. %Note that the user's 3 mobility profile has been compiled by multiple location-based social networks. %, according to the data we extracted from Foursquare. 
The bottom of the figure illustrates the trajectory of the user $u$ that is evaluated, which consists of the following report sequence: Train Station, Cafe, Clothing Store, Bookstore and finally the user enjoys a burger in a Burger Shop. 
We may observe that there is a common pattern among user $u$ and the third trajectory which belongs to user $id_{u'}=2$ who has probably visited a "Mall".
The data analyst computes the social strength of user's $u$ trajectory (up to the 5th report) and the MPs that have been compiled.
The computed social strength 
%MPs that the adversary has in possession, 
reveals, with high probability, that user $u$ is very likely to visit an "Ice-cream Shop" in one of the next moves. Subsequently, after associating those users two users as more highly socially connected, the data analyst can infer the next probable visit of the user in order to provide for e.g. targeted advertisements for "Ice Cream Shops" or related ads.

%% file: approach.tex
\section{$TP^{3}$ approach}
 %closes in line 326
% In this section, we first provide a high-level overview of our approach and then discuss it in detail. 
%% First we provide our design primitives. Then we discuss how we build the Presence Activity Chains utilizing the coresets theory and how the activity chains are exploited to estimate the user's similarity with existing groups of users. 
% Finally, we describe how our approach handles adversary threats and we give the implementation details of our application.

%\subsection{Overview}

In this section we present the %design principles 
main mechanisms of the  $TP^{3}$ system: a) we first build mobility profiles using the theory of \emph{coresets}, that provides a guaranteed approximation sample of the original geolocated data reports stored on-device, 
b) we formulate our \emph{multi-objective optimization problem}, 
%where, given a user trajectory $tr_{u}^{l}$ we compute the social strength of user $u$ with his social ties, in which the user should be alerted and appropriate privacy measures should be taken if a certain privacy threshold is exceeded, 
c) we present \emph{our algorithm for estimating the set of all the non-dominated solutions}, which constitute a Pareto frontier, and d) we apply a set of \emph{serverless privacy operations} at the last stage of the $TP^{3}$ system that aim at minimizing the social strength of the user with any kind of social relationships, ({\it i.e.,} friends or users with similar behavior) to preserve his privacy.

\subsection{Building Mobility Profiles}\label{sec:mobiprof}

%Vana: Dimitri, des to parakatw einai apo to paper "Friendship and Mobility: User Movement In Location-Based Social Networks". Nomizw na grapsoume ena tetoio introduction edw gia na tonisoume (1) people have periodic behavior throughout the day opws ta lene parakatw me ta examples kai (2) oti kanoume model user locations with coresets: coresets is a data reduction technique that allows rapid reduction of the number of geolocated data reports that represent the user MP.

%In particular, we build on the observation that people show strong periodic behavior throughout certain periods of the day alternating between primary (e.g., “home”) and secondary (e.g., “work”) locations on weekdays, and “home” and social network driven locations on weekends. Our model has three components: (1) a model of spatial locations that a user regularly visits, (2) a model of temporal movement between these locations, and (3) a model of movement that is influenced by the ties of the social network (i.e., meeting friends). We capture the temporal dynamics of transitioning between these locations with a day specific periodic transition model. We model user locations with a mixture of Gaussians centered at “home” and “work” locations. On top of this we use a model of “social” movement that governs user behavior over the weekends and weeknights.

$TP^3$ is built based on the observation that people show strong regularities in their behavior while using trajectory-based apps. For example, during weekdays they typically follow specific movement patterns between primary (e.g., "home") and secondary (e.g., "work") locations, while on weekends they present different mobility behavior that alternates between "home" and various social locations. We model the users' movement patterns using \emph{coresets}, a data reduction technique that allows us to significantly reduce the trajectories size while providing a guaranteed approximating sample of the original trajectory stored on-device. 

\noindent
{\bf Theoretical Background.} 
%Vana: olh h parakatw paragraph den prosferei tipota.
%A key challenge we face while building the Presence Activity Chains for groups of users, is the \textit{differentiality} on the manner the user's interact with the system. For instance, there exist users that tend to share their presence activity on a very frequent manner, whereas, others, may prefer to share their presence activity only when reaching major places of interest (such as the restaurant of the university, which is a very known and common venue across university students). Another typical scenario is the one consisting of students and employees in the secretariat of a university department, who have different presence activity chains. Students move by the bus on a specific line and employees utilize their cars to reach the campus. However, the respective activities differ on space and time. Our goal is to evaluate the association of a user to a user group by her behavior over space and time. This is achieved by capturing her closeness to one of these presence activity chains. Therefore, for addressing the aforementioned issue, a sampling technique is required to generate a representative set of points that approximates very well the behavior of each group over space and time and in a certain area. 
%Vana: bres ena introduction giati theloume na kanoume sampling. Mia protash arkei. To parakatw den eimai sigourh oti einai swsto. To user group behavior kanoume approximate me ta coresets h to trajectory tou kathe user?
%Dimitris: Addressed.
We utilize a sampling technique in order to generate a representative set of reports that approximates well the trajectory of a user over space and time rather than keeping all his spatio-temporal reports. In the literature, many different sampling techniques have been proposed such as ~\cite{ohlsson1998sequential}. 
%Vana: ta parakatw einai xarakthristika olwn twn sampling techniques, swsta?
%Dimitris: Me eksairesi to prwto pou anaferoume ws key characteristic ola ta alla einai koina.
%A key characteristic of the sampling technique is that {\bf the sampled data should approximate well the initial trajectory} in space and time. 
The sampling technique should be designed with the following properties: \textit{i) provide {\bf guaranteed approximation} of the initial dataset}, \textit{ii) provide {\bf a sample set of minimal size with bounded loss of information}}, and finally, \textit{iii) have {\bf small algorithmic complexity} that can be executed in memory-constrained environments, such as users' mobile phones in our setting}. In $TP^{3}$'s services, we exploit the theory of coresets\cite{agarwal2005geometric}, which fits our setting and has been recently used to address geometric and graph problems\cite{feldman2017coresets}, such as k-means, k-median, etc. Our approach differs from the above works, since: (1) we consider user trajectories rather than single geospatial reports, and (2) address different privacy goals.
%(compared to approaches such as \cite{feldman2017coresets} where they focus on generating differentially private k-means clusters).
The benefit of coresets is that they constitute a small set which approximates well the original data, and running queries on the coreset produces similar results to the original data. Thus, in our approach: (1) we keep only the reports of the generated coreset when compiling and storing a user's trajectory, (2) this results in significant performance benefits, as the number of reports kept locally on a user's phone is significantly reduced, as extensively evaluated in \cite{Boutsis2016a}.
%Vana: grapse se parenthesi thn diafora sto 2o point parapanw.
%Dimitris: Addressed.

In computational geometry, a coreset $CS$ of a point set $X$ is a sample  that can efficiently approximate the initial set of points $X$. 
% Coresets have the benefit that they comprise a smaller set in terms of size, thus reducing also the space required for storing the data on user's mobile phones, but provide similar results with the original point set. 
%DIMITRIS_PERCOM2018: To apo katw sxoliasmeno einai redundant giati to exoume pei parapanw gia ta coresets.
%Coresets preserve the evolution of a presence activity sequence over space and time, since they approximate the original sequence of points, and therefore fulfill the aforementioned sampling technique's requirements. 
Given a set of user shared spatio-temporal reports $C$, we assume that $C$ can be approximated by a factor $1\pm \epsilon$ from a smaller subset $C^{*}$ of the user's shared spatio-temporal reports. 
More formally,
%C=\gamma * C^{*} + \eta$$ where $\gamma$ is an approximation factor and $\eta$ is an additive error. Additionally, 
for the given point set $C$ and a class of queries $Q$, the following property holds for the coreset $C^{*}$ and for a given $\epsilon$:
$
    (1-\epsilon)Q(C) \leq Q(C^{*}) \leq (1+\epsilon)Q(C)
$

%The above equation illustrates that for a given $\epsilon$ threshold, the coreset approximates the original dataset by a factor $1\pm \epsilon$, which, in our scenario, can be translated to having a set of shared reports that approximate the user's trajectory by a factor $1\pm \epsilon$.

\noindent
{\bf Trajectory Coresets.} 
%We use coresets to generate the states that comprise a PAC. Since it is required to reduce the amount of information stored locally on user's phone, we need to create a sample of the initial trajectory, using a sampling technique, such as coresets. 
%In order to address the adversity of user variety in the way they interact with the crowdsoucing system, w
%Coresets are applied in order to generate a representative sample of the user trajectory in a specific area for all users $u\in \mathcal{U}$ of our system. As aforementioned, coresets allow capturing the behavior of the user in a specific area but with less number of spatio-temporal reports needed.
We build the coreset of the user's $u$ trajectory $tr_{u}^{l}$ by selecting the appropriate spatio-temporal reports that will comprise the coreset as follows:
Without loss of generality, we apply a well-known compression mechanism\cite{Boutsis2016a} on the user's $u$ trajectory to generate a set of spatio-temporal reports that preserve the shape of the user's trajectory but with fewer number of reports. The generated set of reports after applying this scheme is the coreset of the user's trajectory. 
%Vana: den einai xekatharo pws to sample pou les parapanw syndeetai me ta coresets.
%Vana: prwta prepei na peis ti theleis na kaneis capture me to parakatw kai meta poia points tha ginoun included sto coreset.
%Dimitris: Addressed.
% In our system we identify at run-time whether a spatio-temporal report of the user's trajectory should be included or not. This is achieved by calculating whether (and how much) the direction from the initial spatio-temporal report of the sample up to the last one changes above a predefined threshold $\theta$. The algorithm stores the first and last reports and evaluates the inclusion or not of other reports by calculating the percentage change of the tangent or the cotangent. More formally, the algorithm evaluates, for two consecutive reports $p_{u,i}^{\tau}$ \& $p_{u,i+1}^{\tau+1}$, one of the following formulas:
% $    \label{eq:form1}\small
% |\frac{tan(p_{u,i+1}^{\tau+1})-tan(p_{u,i}^{\tau})}{tan(p_{u,i}^{\tau})}|\geq \theta \normalsize
% $
% or
% $
%     \label{eq:form2}\small
%   |\frac{ctan(p_{u,i+1}^{\tau+1})-ctan(p_{u,i}^{\tau})}{ctan(p_{u,i}^{\tau})}|\geq \theta \normalsize
% $.
In our work, we apply a procedure similar to \cite{Boutsis2016a}. This procedure generates a trajectory that preserves the sequence in space and time but approximates the original user trajectory with fewer data reports. In order to decide whether a new trajectory shared by the user can be described by a coreset, we evaluate its reports. Specifically, the algorithm evaluates, for two consecutive reports $p_{u,i}^{\tau}$ \& $p_{u,i+1}^{\tau+1}$, if the percentage change of the tangent is over a predefined threshold $\theta$. More formally,
$    \label{eq:form1}\small
|\frac{tan(p_{u,i+1}^{\tau+1})-tan(p_{u,i}^{\tau})}{tan(p_{u,i}^{\tau})}|\geq \theta \normalsize
$.
High values of $\theta$ ($\theta > 0.0005$) define a stricter sampling (and a smaller size of coreset) compared to values close to zero.
%For this purpose, we define an appropriate distance function.

%Vananew: Dimitri, den einai xekatharo giati kanoume to parapanw bhma kai giati to parakatw. Pote apofasizoume poio report na baloume sto dataset? sto parapanw bhma h sto parakatw? Mporeis na to exhghseis ligo kalytera? 
%Dimitris: Addressed.  Auto pou theloume na poume einai oti apofasizoume ean ena report tha mpei sto coreset i oxi vlepontas kata poso ta idi yparxonta reports mporoun na einai to idio antiproswpeutika. Diladi, an yparxei kapoio existing report sto coreset pou einai poly kontino se auto pou theloume na prosthesoume opote kai den tha to prosthesoume.
\noindent

%DIM_ICDCS_2021: to evala se comment. prepei na to meiwsoume 
% {\bf Distance Function. } To decide whether a new spatio-temporal report belongs to the coreset of the user's trajectory we first compute its distance from the spatio-temporal reports that comprise the coreset. 
% We define this distance function as:
% $    dis(p_{u,i}^{\tau},CS)=min_{cs\in CS}|p_{u,i}^{\tau}-cs|
% $  
% where $|\bullet|$ is a distance metric, where a distance function such as the Haversine distance, the Euclidean distance or the Chebyshev distance can be utilized. 
% %Vana: bale references gia ola ta distance functions.
% %Dimitris: Addressed.
% %One of the key requirements of our distance function is to compute this in minimum computational time, in order to keep the computational load on the mobile devices small. 
% We utilize the Chebyshev distance as the distance metric, which has smaller computational overhead on a mobile device compared to the Euclidean distance\cite{feldman2017coresets}.
% % and for any two points $(x_1,y_1)$ and $(x_2,y_2)$ is defined as 
% % \begin{align}
% %   L_{\infty}=max(|x_2-x_1|,|y_2-y_1|)  
% % \end{align}
% If the value of $dis(p_{u,i}^{\tau},CS)$ is below a certain distance $R$, then the specific spatio-temporal report $p_{u,i}^{\tau}$ belongs in the coreset $CS$ of the trajectory. 

\subsection{ Our multi-objective optimization problem}
\subsubsection{Social Strength Minimization}
Assume a user trajectory $tr_{u}^{l}$ comprising a list of data reports shared by user $u$. 
Then assume that a user wishes to share a new trajectory. %, which will be added as a new part of the existing trajectory.
The question is whether it is safe for the user to issue a query sharing this trajectory.
The role of $TP^3$ is to evaluate the safety for the user to issue it and then apply appropriate privacy measures. 
Thus, given a set $\mathcal{B}$ of MPs $\mathcal{G}_{u_{k}}$ that belong to possible social user ties and a threshold $\delta$, we compute the social strength of the user's trajectory $tr_{u}^{l}$, compared to the users represented by the MPs, using a score function as follows:
%Vana2: Dimitri, xereis to trajectory tou tr? edw den theloume na kanoume alert ton user? prepei na grapseis kalytera to scenario.. assume a user with a tr that has a list of points kai thelei na kanei post ena data report san part of the trajectory. the question is whether it is safe to allow this report or alert the user that he may compromise his privacy. 
%Vana2: isws olo to section C na prepei na paei meta san section III.B gia na einai xekatharo ola ta steps toy approach mas. ti les? 
%Dimitris: To mono mou concern einai oti den prepei to minimization problem na to exoume sto model? Tha to ksanakanw rewrite tote opws mou grapsate to scenario. An paei meta to section III.B, pisteuete oti tha einai kalytero to flow? Apo tin alli an meinei mono to attack mazi(Social Link exploitation attack) me to example tha einai wraio. (me mia mikri allagi mono sta figure mas gia na exei swsti roh.).
%Dimitris: Na to metaferw kai na to kanw rewrite to scenario?
% i.e. it computes the probability $\mathcal{P}(tr_{u_{1}}^{l_{z}}|tr_{u_{2}}^{l_{y}})$ (equation \ref{eq:prob}) for each trajectory in every MP. 
% \begin{align}\label{eq:scorefun}
$  score(tr_{u}^{l},\mathcal{B})=\frac{1}{|\mathcal{B}|}\cdot \sum_{\forall \mathcal{G}_{u_{k}}\in \mathcal{B}}\mathcal{S}(tr_{u}^{l},\mathcal{G}_{u_{k}})
$

\subsubsection{Performance Maximization}
The second metric we consider in our multi-objective problem is the requests success rate(RSR). The RSR has been introduced in recent works \cite{palade2019evaluation} as a performance metric of serverless functions. For a given memory allocation $m\in M$ from a set $M$ of possible memory allocations, the requests success rate $\lambda_{m}$ is defined as the ratio of the number of user requests successfully served by this memory allocation $m$, $sucreq_{m}$, to the overall number of the requests submitted by the users to the system, $total_{m}$. In our work, our goal is to maximize the execution performance for all possible given memory allocations $m\in M$. Thus, our objective can be formulated as follows:
$
    EP(m) = \max (\lambda_{m})=\max (\frac{\# sucreq_{m}}{\#total_{m}}),\forall m\in M
$.

\subsubsection{Spending Budget minimization}
The third metric we consider in our multi-objective problem is the spending budget $SB$. To compute this metric, we applied a pricing model similar to the one used by popular cloud providers like IBM (https://cloud.ibm.com/functions/learn/pricing). The metric considers a basic rate $c_{r}$ which denotes the amount of monetary units to pay per GB of data per sec, the average execution time of the serverless privacy preserving operation $avgT_{\mathcal{F}(\bullet)}$, the memory allocation $m\in M$ allocated for the execution of the function and the number of successful requests served by the system, $sucreq_{m}$. More formally,
$
    SB_{m}= c_{r} \cdot avgT_{\mathcal{F}(\bullet)} \cdot m \cdot sucreq_{m}
$.

Thus, for a finite cloud operator budget $\mathcal{C}_{b}$ monetary units, our goal is to maximize the difference $\mathcal{W}(m)$ between $\mathcal{C}_{b}$ and $SB_{m}$. More formally,
$
    \mathcal{W}(m)= \mathcal{C}_{b} - SB_{m}
$.

\noindent
{\bf Problem Definition.} More formally, our problem can be formulated as a maximization problem as follows:
%\vspace{-1em}
\begin{align}
\max \mathcal{T}(EP(m),\mathcal{W}(m))\\  
s.t.\min \ score(\mathcal{F}(tr_{u}^{l}),\mathcal{B}) < \delta \\
\mathcal{W}(m) > 0
\end{align}
where $\mathcal{T}(\bullet,\bullet)$ is our objective function that considers both the execution performance and the spending budget.

\subsection{Pareto Frontier Search Algorithm}

% Utilizing the serverless model, we aim at choosing the appropriate memory allocation that will enable us to accommodate a set of privacy operations that minimize the social strength between users of trajectory analytics applications, while achieving high rate of requests and minimizing the spending budget. 
In $TP^{3}$, we solve a multi-objective optimization problem where we aim to balance the trade-off between the performance maximization and the required budget, while preserving privacy for the user trajectories. One of the most common ways of detecting appropriate solutions in such problems is constructing the Pareto frontier. In order to detect the optimal solutions in the examining search space, we need to define the notion of dominance \cite{ben2012expert}. Given two memory allocations $m_{1}$ and $m_{2}$, $m_{2}$ \textit{dominates} $m_{1}$($m_{2}\succeq m_{1}$) if one of the following criteria is met: (1) the spending budget for $m_{2}$ is less than equal than the one required for $m_{1}$ and the performance of $m_{2}$ is greater than equal to the performance of $m_{1}$ or (2) $m_{2}$ requires strictly smaller budget than $m_{1}$ and also the performance of $m_{2}$ is greater than or equal to the performance of $m_{1}$.
However, computing the Pareto frontier is a computationally costly process. A naive approach is to enumerate all possible combinations of memory allocations, performance the spending budget.  Such an exhaustive search algorithm has exponential complexity $O(m^{|\mathcal{K}|})$ as it generates all these $m^{|\mathcal{K}|}$ possible allocations. We propose a novel approach that detects near-optimal memory allocations in an efficient and fast way without enumerating all the solutions. Our greedy algorithm approximates the Pareto-optimal frontier by selecting the appropriate memory allocation for the serverless privacy function that is affected the most in its performance by memory allocation. Starting with the memory allocation that helps maximize the performance, we traverse the frontier to select the most appropriate one that minimizes the spending budget. By doing so, it is not required to enumerate all possible solutions.

\subsection{Serverless Privacy-Preserving Operations $\mathcal{F}(\bullet)$}

$TP^{3}$ aims for social strength minimization against a set of MPs that it has already compiled and appropriately stored. 
%Despite the large number of approaches related to privacy in location-based services have been proposed in the literature \cite{yang2016privcheck,ToGS14}, existing works attempt to protect users from inference attacks arising from the publication of their shared crowdsourcing data by focusing on single users. On the other hand, in our scenario, we focus on a totally different problem, which is alerting users when their publicly shared data are adequate to associate them with groups of users. Therefore, these works do not fit in our setting. The goal is to reduce the probability a user can be associated to a user group. 
This is achieved by applying privacy-preserving operations that distort the users' trajectories and therefore minimize the social strength with any social ties. Such techniques include \textit{\bf spatial-location cloaking} approaches \cite{SiksnysTSY10}, \textit{\bf temporal cloaking} methods \cite{Gruteser2003}, addition of redundant \textit{\bf dummy locations} \cite{Cho2009} and \textit{{\bf path confusion}} techniques \cite{Hoh2010}. 
%
%{\bf Privacy-Preserving Operations. } As aforementioned, in order to minimize the social strength, PaROUSiA applies one of the following privacy-preserving operations each time the user is about to issue a new spatio-temporal data report and the user's trajectory surpasses a certain similarity threshold when compared to the compiled MPs. 
%
Depending on the user's required level for privacy, $TP^{3}$ applies the appropriate privacy operation each time the user wishes to publish new trajectory data to minimize the social strength below the $\delta$-threshold.
%Vananew: Dimitri, allaxa ligo thn parapanw protash, des thn.
%Dimitris: To allaksa giati den evgaze sense opws to diavaza. Opws itan grammeno legame oti kathe fora mporei px na dialeksei metaksy twn diaforetikwn privacy operations, enw ayto pou kanoume einai gia diaforetiko typo application kai gia to diaforetiko level of privacy kanei apply mia sygkekrimeni. To idio grafoume antistoixa kai einai reflected sto implementation.

\noindent{\bf Cloaking. } In the \textit{Spatial-location cloaking} privacy model \cite{SiksnysTSY10}, the exact location of the user is replaced by a broader spatial region termed \textit{cloaking region} (CR). 
%CR is defined as an arbitrary spatial region. Specifically, 
This privacy-preserving operation takes as input a spatio-temporal data report and returns a region cell $\widehat{rc}_{p_{u,i}^{\tau}}$ 
(rather than the exact spatio-temporal location)
that includes the spatio-temporal data report $p_{u,i}^{\tau}$ the user wants to publish. 
% More formally, 
% $\widehat{rc}_{p_{i}^{\tau}}=(p_{u,i}^{\tau}.lat\pm \vartheta,p_{u,i}^{\tau}.lon\pm \vartheta)$
% where $\vartheta = (d+d(2\sqrt{2}-1))$ and $d$ is the width of the grid cell. 
This technique simply blurs a user spatio-temporal report into an uncertainty region.
%which stratifies the user's specified minimum spatial area requirement.
%Vanaicd: Dimitri, koitaxe an h parapanw protash poy ebala einai swsth.
%Dimitris: Nai, einai swsti giati ousiastika kanei blur mesa sto area to report alla at the same time kanei satisfy to minimum requirement.
%ensures that the user's actual location is known to the system not more precisely than the cloaking region.
%
%The privacy operation takes as input the generated list of tupples with similarities to user groups from the previous step of the algorithm, the user's trajectory and the MPs $\mathcal{B}$. 
%For efficiency purposes, we do not re-evaluate the similarity against all MNPs but only with those that the user has similarity within a predefined threshold $\delta$.  
% Instead of publishing the actual geospatial report $p_{u,i}^{\tau}$, $TP^{3}$ publishes four geospatial reports that define the region cell $\widehat{rc}_{p_{u,i}^{\tau}}$.
A larger region size indicates a more strict privacy requirement, at the expense of not providing useful information for the system.
%Vanaicd: Dimitri, des kai thn parapanw protash, ayto itheles na peis?
%Dimitris: Nai, oso megalytero einai to region toso pio privacy concerned einai o xristis alla me mikrotero utility gia ton idio en telei.
%For example, in this privacy operation, we consider as more useful those region cells where the area covered is below a certain threshold (a region cell of size 100m x 100m is more useful compared to one of size 1000m x 1000m).

% \begin{figure}[t!]\centering
% \begin{minipage}{\linewidth}\centering
% \includegraphics[width=\linewidth]{dataset/flow.png}
% \caption{Flow of $TP^{3}$.}\label{fig:flow}
% \end{minipage}\hfill
% \end{figure}

% \begin{figure*}[t!] \centering
% \begin{minipage}[b]{0.2\linewidth}
% \centering
% \includegraphics[width=0.65\textwidth]{mapsActivity.png}
% \caption{$TP^{3}$'s Alert message on report sharing.}\label{fig:app1}
% \end{minipage}\hfill
% \begin{minipage}[b]{0.2\linewidth}\centering
% \centering
% \includegraphics[width=0.65\textwidth]{settingsActivity2.png}
% \caption{$TP^{3}$'s Privacy Settings Configuration Page}\label{fig:app2}
% \end{minipage}\hfill
% \begin{minipage}[b]{0.55\linewidth}
%      \centering
%     \includegraphics[width=1.05\linewidth]{implementation.png}
%     \caption{High-level operation overview of $TP^{3}$}
%     \label{fig:implementation}
% \end{minipage}
% \end{figure*}

\noindent {\bf TempCloaking. } Compared to the Spatial Cloaking model, this privacy-preserving model \cite{Gruteser2003} uses time transformation and delays the user's response by a time period. That is, for two consecutive time instances $\tau_{1}, \tau_{2}$, the time instance $\tau_{2}$ of a new report $\widehat{p_{u,i}^{\tau_{2}}}$ is set to the time $\tau_{1}$ plus a random cloaking factor. 
% More formally,
% $\widehat{p_{u,i}^{\tau_{2}}}=p_{u,i}^{\tau_{1}+rf},\ where\ rf = randomFactor(\tau_{1},\tau_{2})$
% where $randomFactor(\tau_{1},\tau_{2})$ returns a random timewindow to add to $\tau_{1}$. 
In $TP^{3}$, the timestamp value of the trajectory's data reports is changed accordingly by a specific amount of time, which consequently leads to a different trajectory, thus reducing the similarity with the compiled MPs.  %(as we further show in our experimental evaluation). 

% \begin{figure*}

% \end{figure*}

\noindent {\bf Dummy Locations. } An alternative approach to applying time or spatial transformations is this privacy-preserving model \cite{Cho2009} where a number of \textit{dummy locations} is generated. The user, instead of reporting the actual location, reports one or more locations which are very close to the actual one. Thus, for a given spatio-temporal data report $p_{u,i}^{\tau}$, a list of one or more dummy spatio-temporal data reports $\widehat{p_{u,ii}^{\tau}}$ is generated. 
% More formally, 
% $\{\widehat{p_{u,i1}^{\tau}},\widehat{p_{u,i2}^{\tau}},...,\widehat{p_{u,iN}^{\tau}}\}=\{\widehat{p_{u,ii}^{\tau}} : p_{u,i}^{\tau}+\mathcal{N}(\mu,\sigma)\}  
% $
% where $\mathcal{N}(\mu,\sigma)$ is a Gaussian noise with mean value $\mu$ and deviation $\sigma$. 
Instead of publishing the trajectory with only the actual spatio-temporal reports $p_{u,i}^{\tau}$, $TP^{3}$ publishes the list of the dummy spatio-temporal reports generated, including the original ones. 
%Here, we consider as useful the spatio-temporal reports that reside within a radius $w$ from the actual spatio-temporal report.
%The larger the distance from the actual spatio-temporal report, the smaller its usefulness to the system.
%region size indicates a more strict privacy requirement

\noindent {\bf Path Confusion. } This privacy model \cite{Hoh2010} differs from the previous models, since perturbations of the previous locations are applied in order to obfuscate and reduce the similarity with the MPs. Given a set of spatio-temporal reports $p_{u,i}^{\tau}$ the user wants to share, the goal is to apply a perturbation technique $pert()$ that changes the actual trajectory of the user (i.e. publish another report instead of the actual one), which results to a different user trajectory. The perturbation process considers up to $q$ sequential spatio-temporal reports to perturb and changes their sequence.
%Vanaicd: einai sequential ta reports, swsta?
%Dimitris: Nai, koitaei ta sequentials px se emas an dyo sequential kai genika ana q sequential.
% More formally, 
% $\{\widehat{p_{u,i1}^{\tau}},\widehat{p_{u,i2}^{\tau}},...,\widehat{p_{u,iN}^{\tau}}\}=pert(\{p_{u,i1}^{\tau},...,p_{u,iN}^{\tau}\},q)  
% $.
\begin{figure}[t!]\centering
\begin{minipage}{0.7\linewidth}\centering
\includegraphics[width=\linewidth]{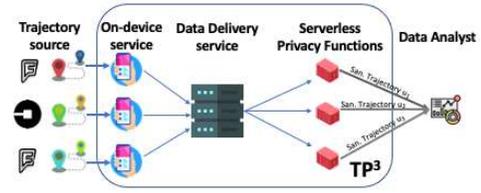}
\caption{Architecture Overview.}\label{fig:architecture}
\end{minipage}\hfill
\end{figure}

%\subsection{Architecture}

\section{Implementation}

%In this section we describe the architecture of $TP^{3}$. 
%An overview of the flow in our LPPS can be found in Figure \ref{fig:flow}. 
%, where we illustrate how our approach evolves from the initial report up to the disclosure to the third party application.

\begin{figure*}[htp]
\begin{minipage}{0.215\linewidth}
\centering
\includegraphics[width=0.58\linewidth,angle=270]{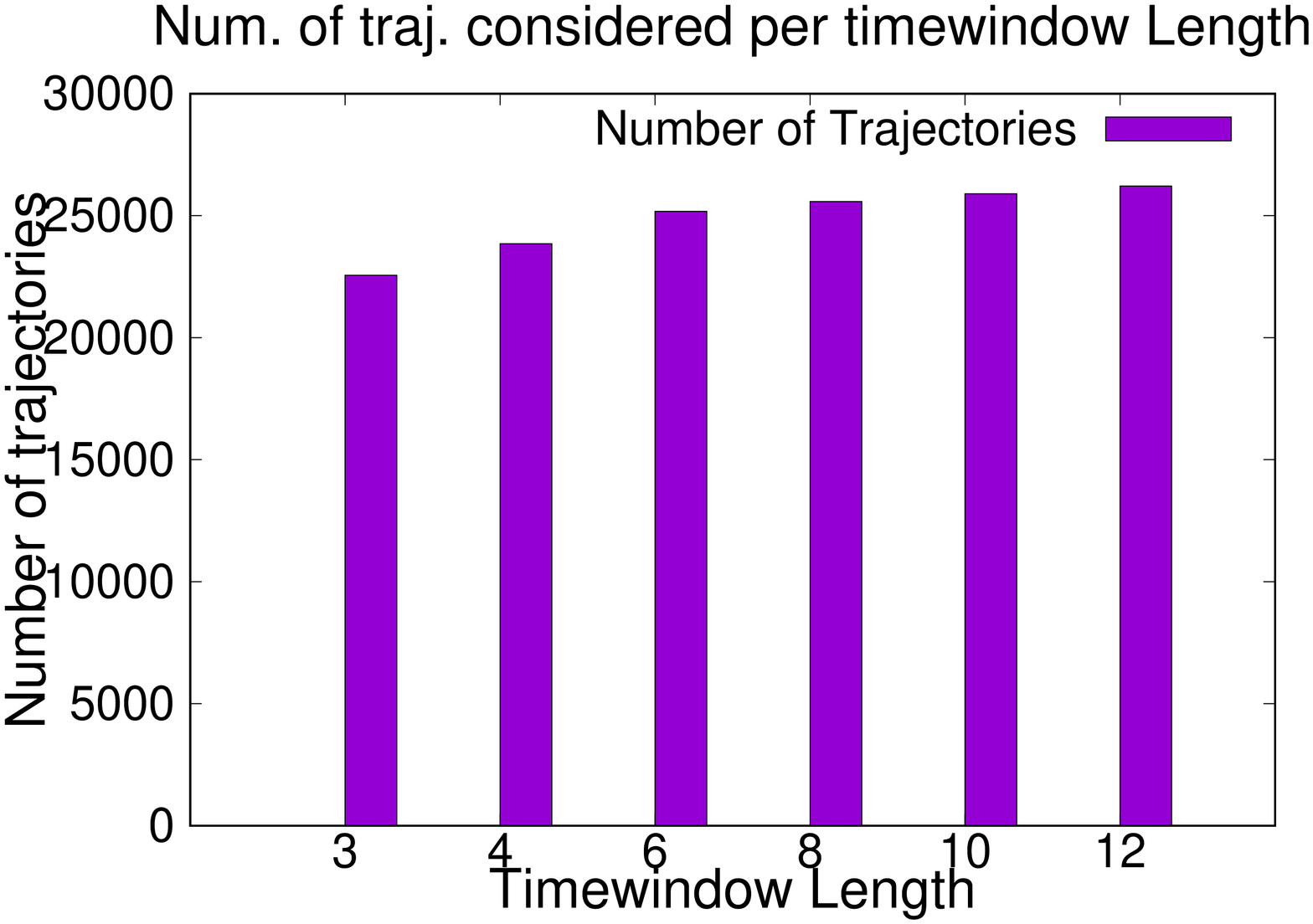}
\caption{Num. of considered  trajectories per timewindow}\label{fig:considertraj}
\end{minipage}\hfill
\begin{minipage}{0.215\linewidth}
\centering
\includegraphics[width=0.58\linewidth,angle=270]{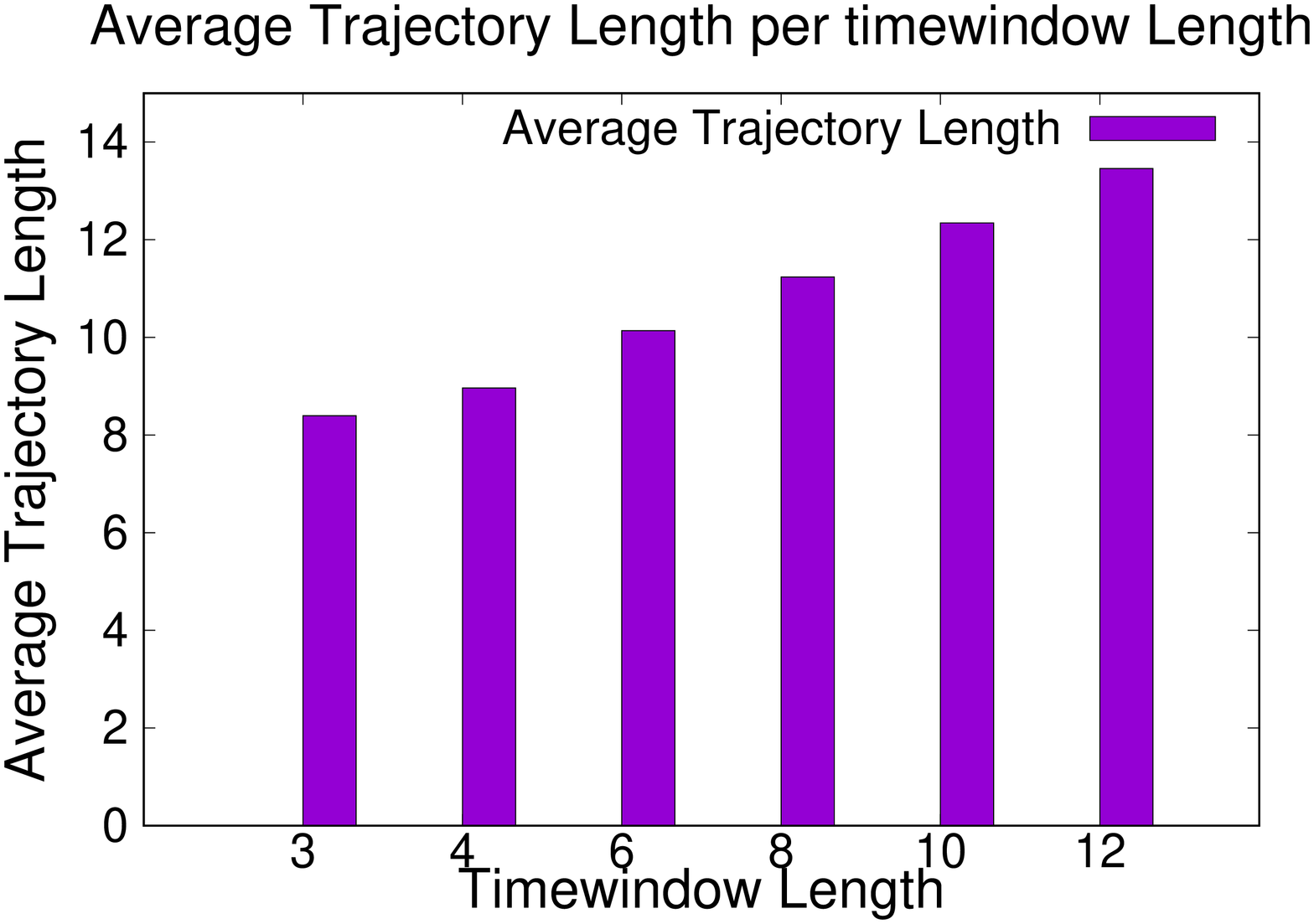}
\caption{Average length of trajectories considered}\label{fig:avgchain}
\end{minipage}\hfill
\begin{minipage}{0.215\linewidth}
\centering
\includegraphics[width=0.87\linewidth]{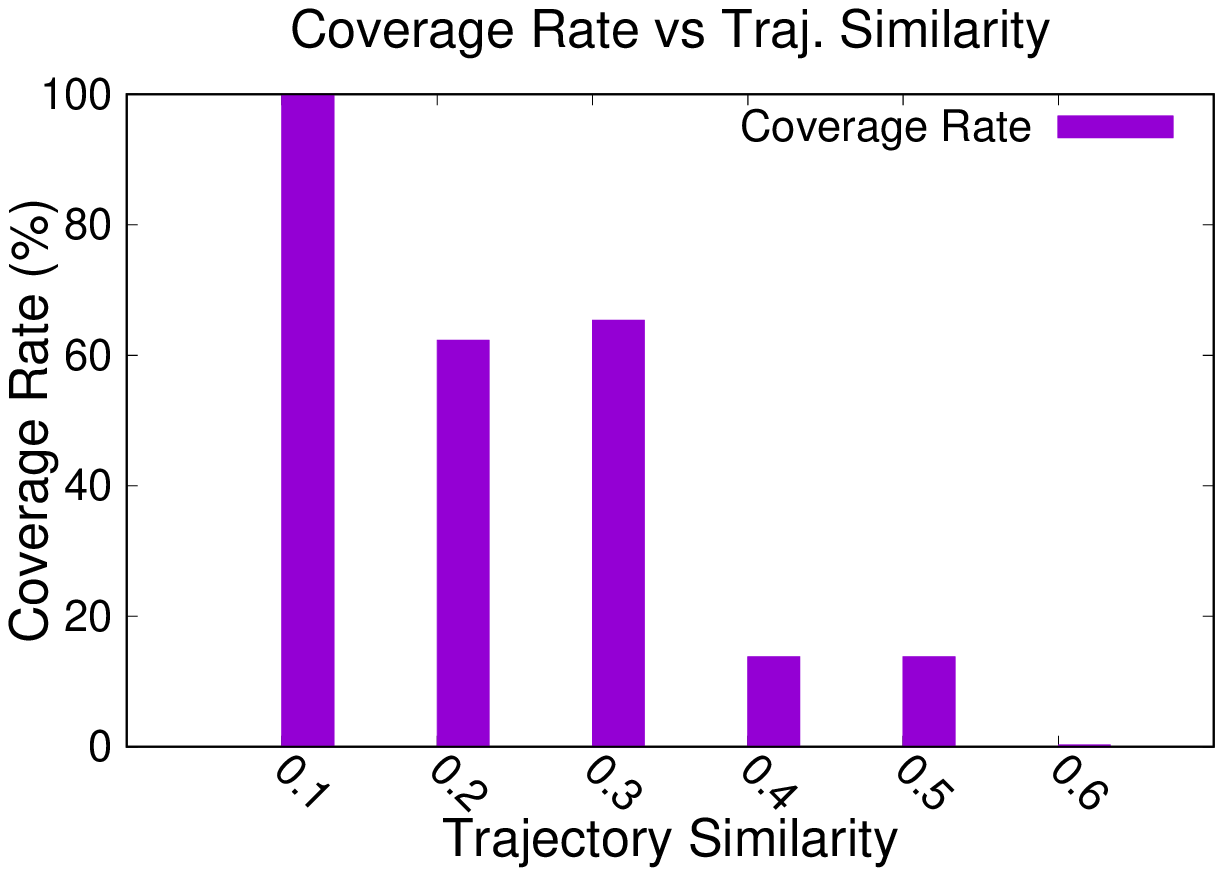}
\caption{Cov. Rate vs Traj. Similarity}\label{fig:covRate}
\end{minipage}\hfill
\begin{minipage}{0.215\linewidth}
\centering
\includegraphics[width=0.87\linewidth]{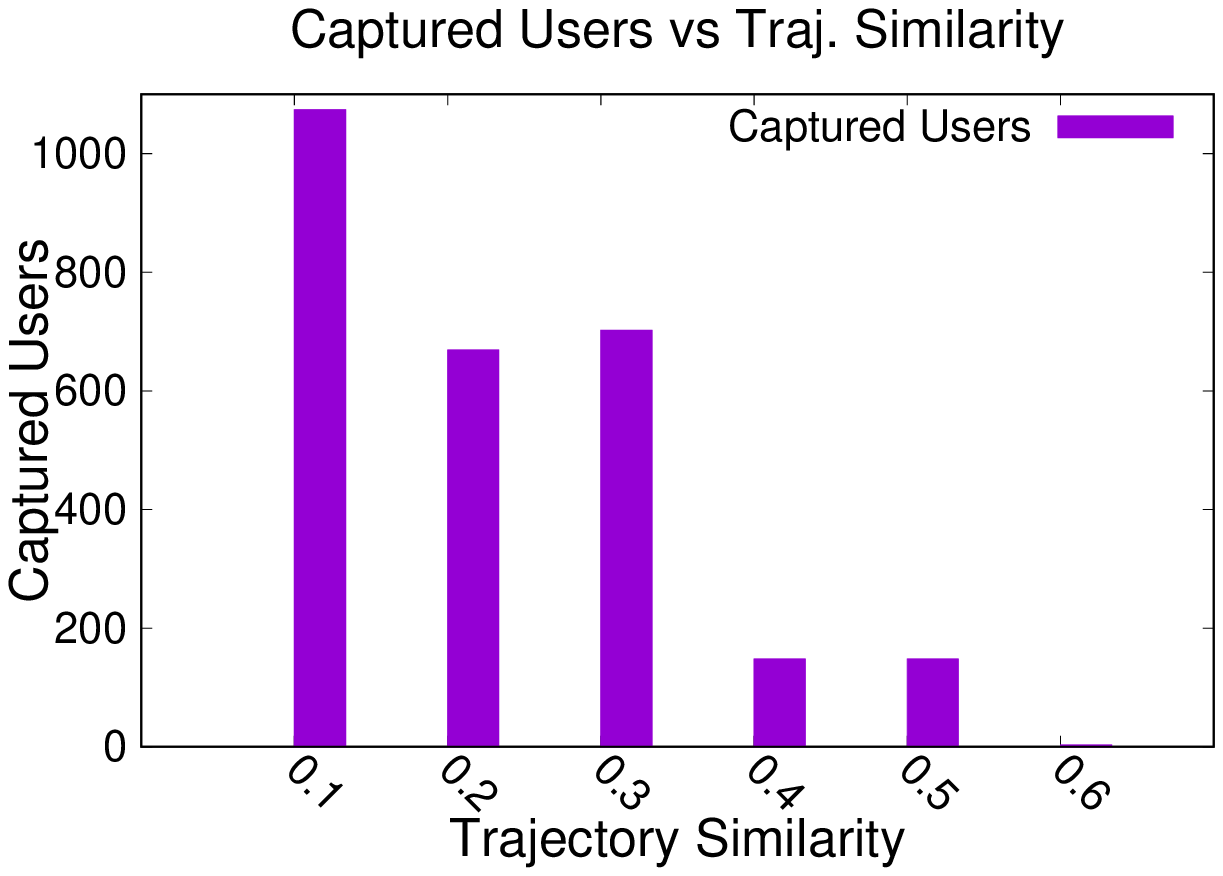}
\caption{Captured Users By MPs.}\label{fig:covRateAbs}
\end{minipage}
\end{figure*}

\noindent
{\bf On-device $TP^{3}$ service.} For the on-device service of $TP^{3}$ on the  Android devices we utilized the Android Development Framework. We implemented an Android Service that works in concert with the Waze (https://www.waze.com/) and  CrowdAlert (http://crowdalert.aueb.gr) and allows for the secure sharing of trajectories. Currently, it provides an API to those apps from which calls to trajectory-publishing methods are redirected through the Data Delivery Service of $TP^{3}$.
If the user is willing to share a trajectory or a list of past trajectories with a data analyst, the Android Service of $TP^{3}$ system is triggered and prepares appropriately the trajectory or the list of trajectories for delivering through the Data Delivery Service of $TP^{3}$, in order to be evaluated based on his personal privacy preferences.
% The Android Service extracts information from his locally stored social graph, in order to first evaluate the user's similarity against his social ties. 

% For this purpose, it asks the Data Ingest service of $TP^{3}$ for the mobility profiles of the  specific subset of users (his social ties) to perform that comparison locally. Upon finishing, it forwards the report to the system for further evaluation by the Social Link Alert service as described below.
% In Figure \ref{fig:app1}, we see a snapshot of this Android Service, when a specific report on the Foursquare location-based service can associate the user with a mobility group. Finally, in Figure \ref{fig:app2}, we illustrate the privacy settings configuration activity, where the user is able to select between levels of privacy, without needing to adjust by his own any threshold values. 

%DIM_ICDCS_2021: Mixali, prosthese to content gia to kafka edw.
{\bf Data Delivery service.} The Data Delivery service in $TP^{3}$ is responsible for processing trajectory shared from the Android Service. The Data Delivery service employs Apache Kafka as its fundamental building block. Apache Kafka is one of the most popular pub/sub systems and it is used for propagating millions of messages per second between a set of producers and consumers or across different services. Apache Kafka, apart from its ultra high throughput, uses replication, thus guarantying  zero lost messages. Each user's on-device service acts as a Kafka Producer and sends trajectories in specific topics, based to the user's personal privacy preferences. Kafka Consumers constantly poll trajectories from these topics and send them through appropriate HTTP endpoints to the OpenFaaS Gateway to invoke the respective Serverless Privacy Function. Afterwards, the sanitized output is provided through appropriate HTTP endpoints to the data analyst. 

{\bf Serverless Privacy Functions.} This component sanitizes the user's trajectory before being publishing it to the data analyst. $TP^{3}$ comprises three different modes of privacy models (loose, moderate and strict), in which different privacy operations are applied in order to distort the users' trajectories. The \textit{loose privacy mode} corresponds to the application of the Dummy Location or Path Confusion privacy function, depending on the type of trajectory app ({\it i.e.,} Path Confusion is suitable for applications where the perturbation of the reports in a trajectory is preferred rather than suppressing them, such as venue recommendation applications), which is encapsulated in the message received from the on-device $TP^3$ service. 
The \textit{moderate privacy mode} corresponds to the Cloaking function and finally, the \textit{strict privacy mode} refers to the TempCloaking function, which changes the nature of the user's trajectory. Finally, the sanitized trajectory is forwarded through the appropriate HTTP endpoints to the data analyst, preserving user's privacy and utility. Our architecture overview is given in Figure \ref{fig:architecture}.

%An overview of our implementation details can be found in Figure \ref{fig:implementation}, where we provide a basic overview of our system and discuss the respective services. We illustrate how these services are interconnected to allow generated reports from the users' location-based apps being captured from the on-device service, being processed by the Data Ingest service, being stored by the Mobile Profile Storage service, and finally, being published as sanitized trajectories to the third party app.
%Vana_1: sthn parapanw protash.. ti shmainei "go through the .." grapse ti ginetai se ayta ta components.. 
%Dimitris: Addressed.

%% file: experiments.tex
\section{Experimental Evaluation}\label{sec:evaluation}

\begin{figure*}[htp]
\begin{minipage}{\linewidth}
\centering
\subfigure[Path Confusion]{
\includegraphics[width=0.215\linewidth]{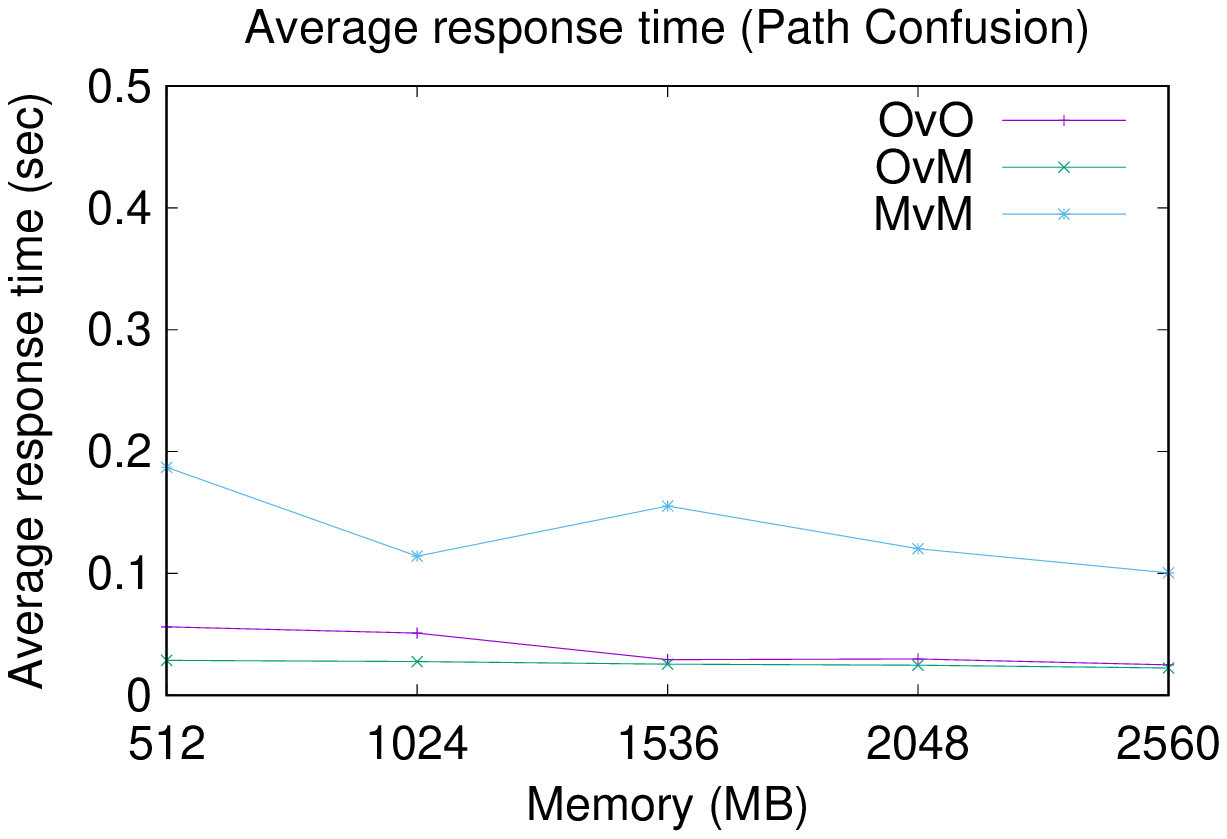}\label{fig:pathconfavgrec}}
\hfill
\subfigure[Dummy Locations]{
\includegraphics[width=0.215\linewidth]{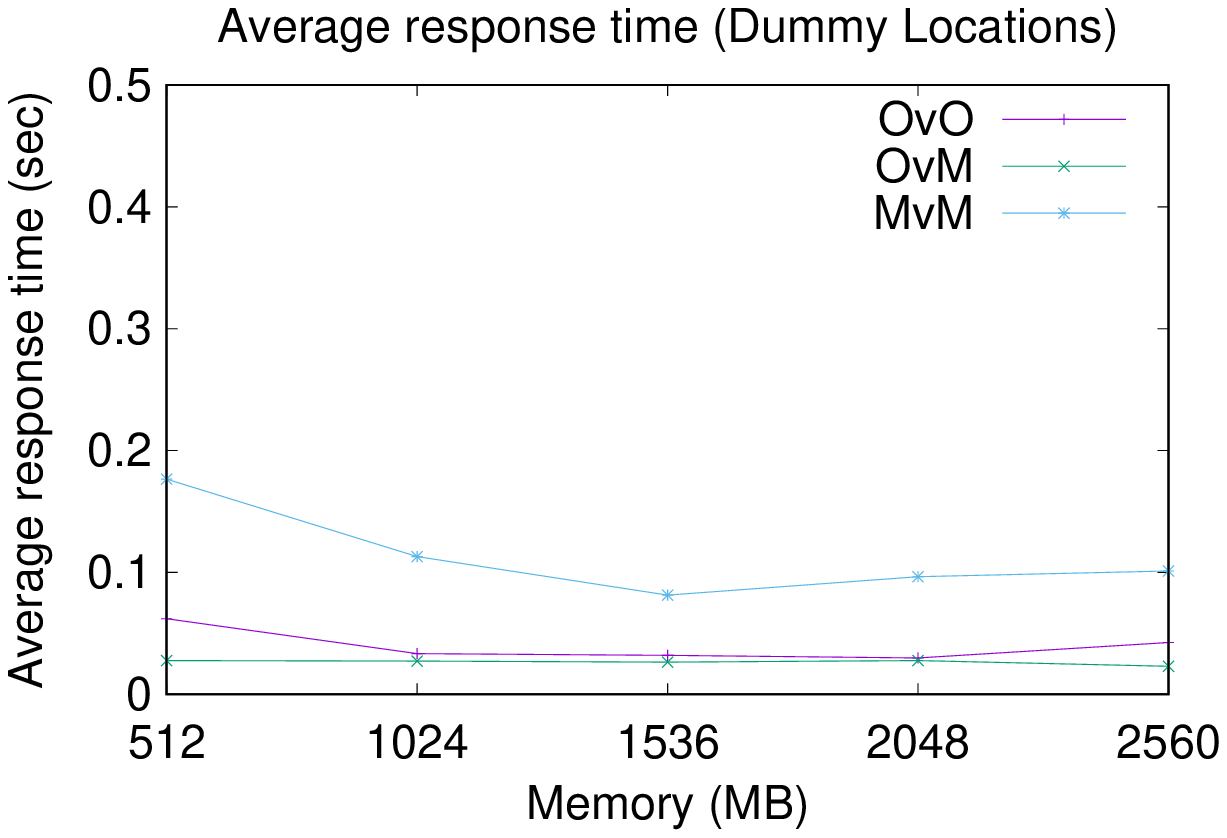}\label{fig:dummyavgrec}}
\hfill
\subfigure[TempCloaking]{
\includegraphics[width=0.215\linewidth]{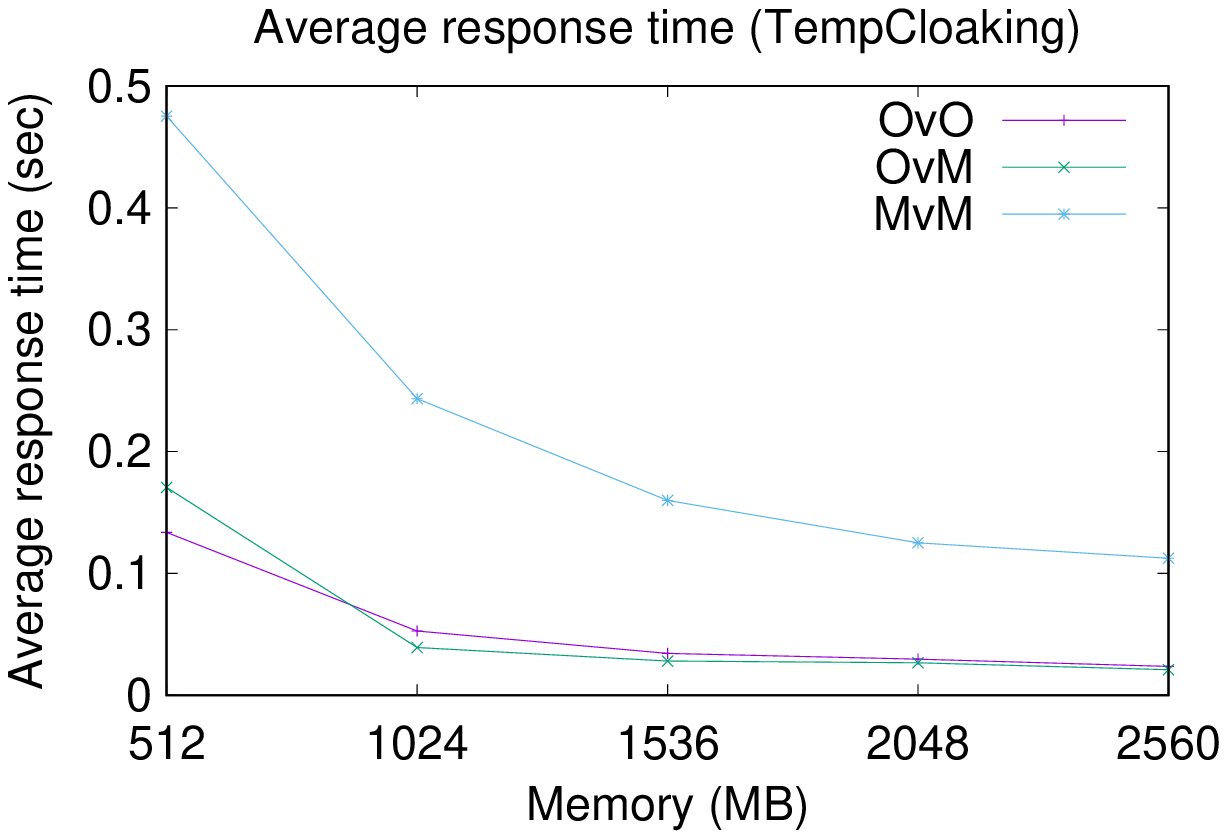}\label{fig:tempcloakavgrec}}
\hfill
\subfigure[Cloaking]{
\includegraphics[width=0.215\linewidth]{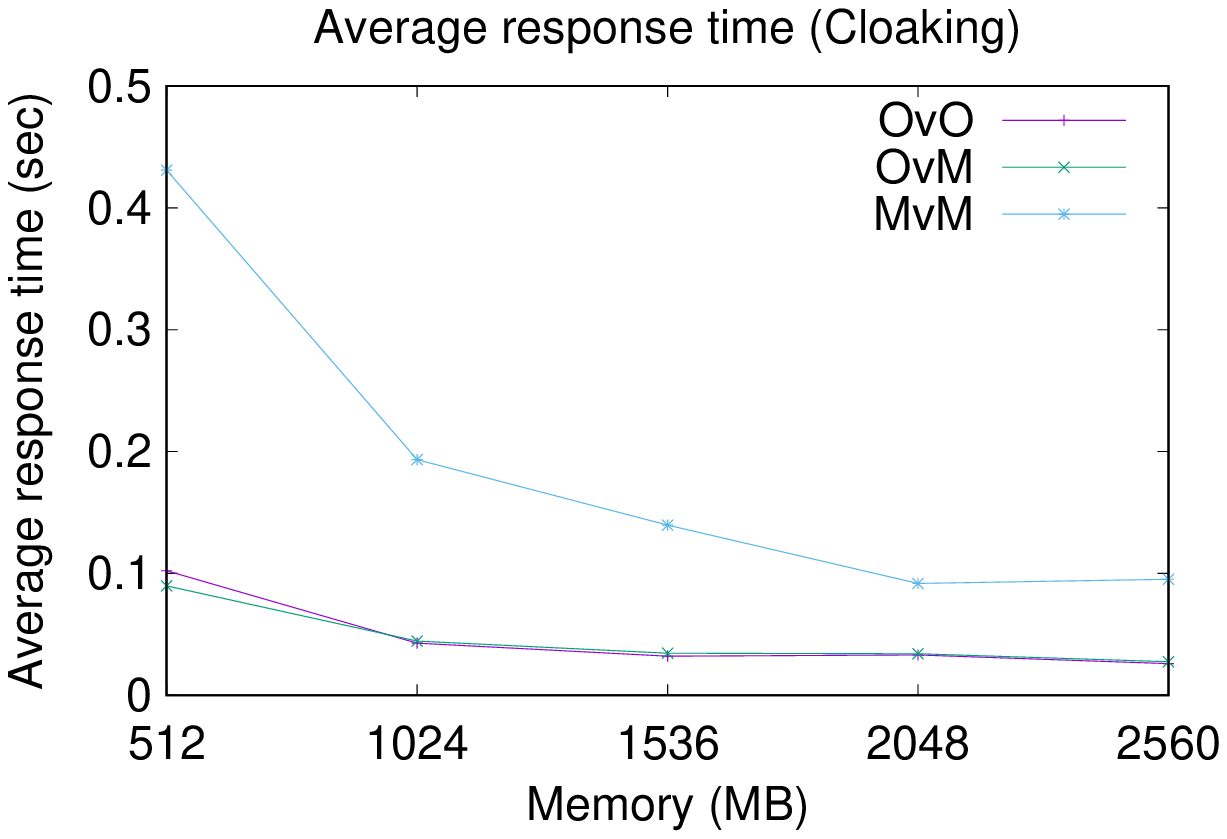}\label{fig:spatcloakavgrec}}
\hfill
\caption{Average response time per privacy operation}
\end{minipage}
\end{figure*}

We conducted series of experiments to identify the effect of different parameters on the efficiency of each privacy operation in minimizing the social strength among the users and to show the benefits of employing a serverless model to maximize the system's performance and minimize the cost. 
%and the benefit of our approach that considers trajectories. Specifically, w
We answer the following questions: 1) how memory allocations affect the average response time,
%and which one could provide a better minimization of the user's social strength, 
2) how the memory allocations affect the throughput in terms of request/sec in each privacy operation, 3) how the different memory allocations affect the RSR, 4) what is the effect of the trajectories' length to the social strength minimization, 5) how the stored report size is affected by each privacy operation, 6) how the memory allocations affect the spending budget of the system operator, 7)  what is the  effect of the percentage of trajectories that the data analyst has in his possession, 8) how the percentage of the sanitized trajectories affects the utility and finally 9) how $TP^{3}$ performs compared to state-of-the-art techniques.

\subsection{Experimental Setting}

\begin{figure*}[htp]
\begin{minipage}{\linewidth}
\centering
\subfigure[Path Confusion]{
\includegraphics[width=0.215\linewidth]{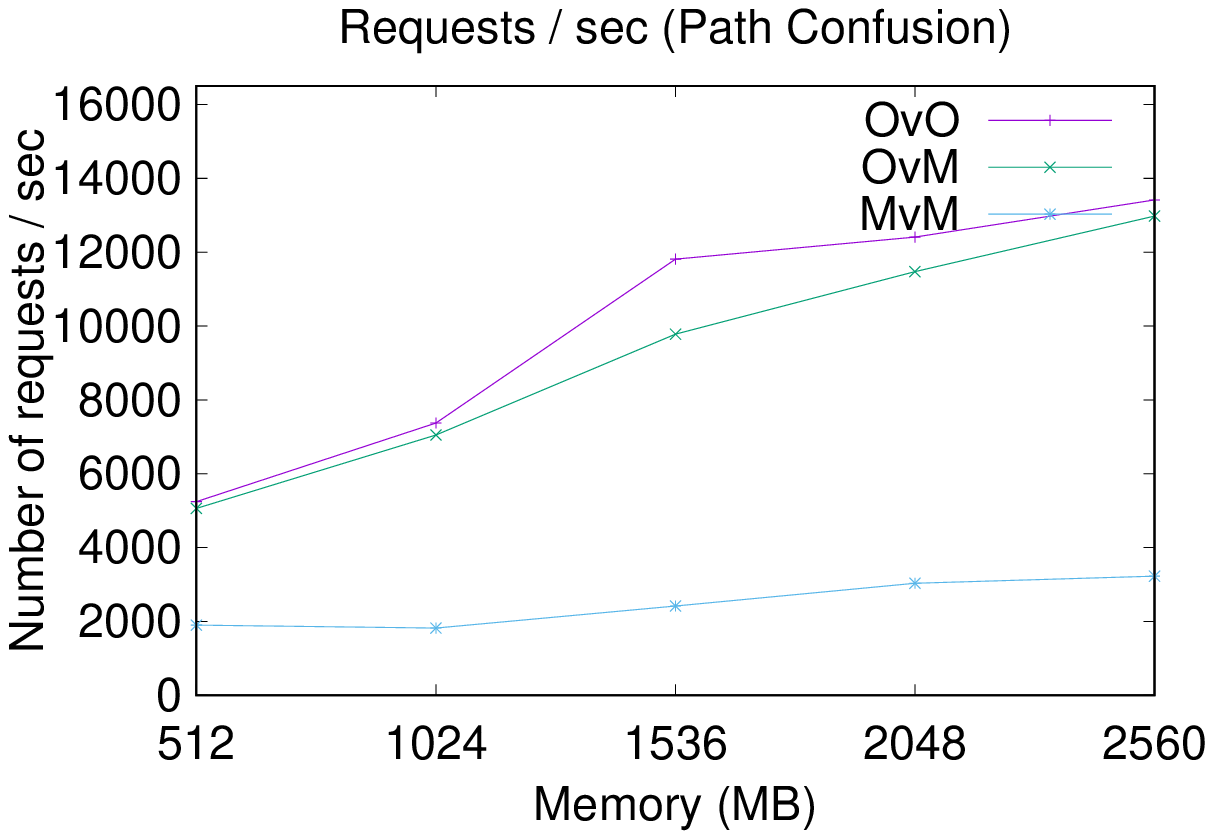}\label{fig:pathconreqsec}}
\hfill
\subfigure[Dummy Locations]{
\includegraphics[width=0.215\linewidth]{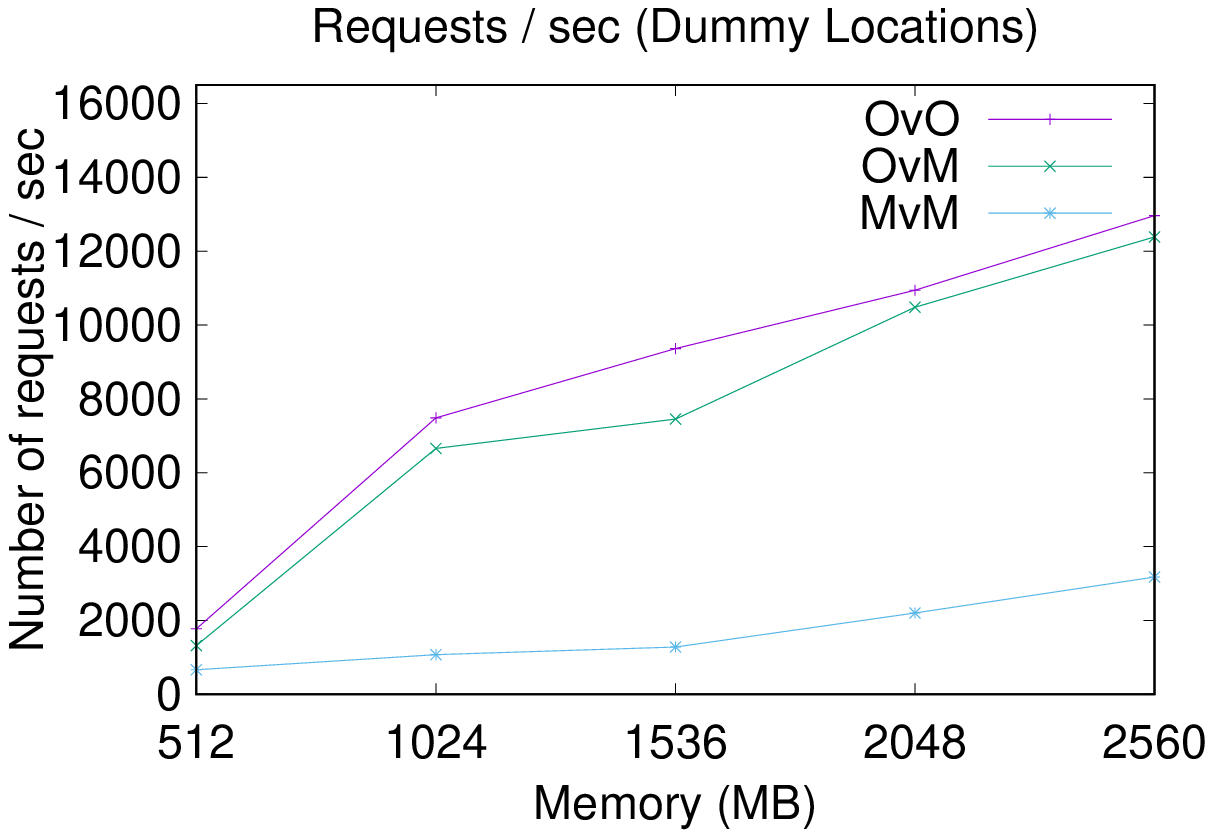}\label{fig:dummyreqsec}}
\hfill
\subfigure[TempCloaking]{
\includegraphics[width=0.215\linewidth]{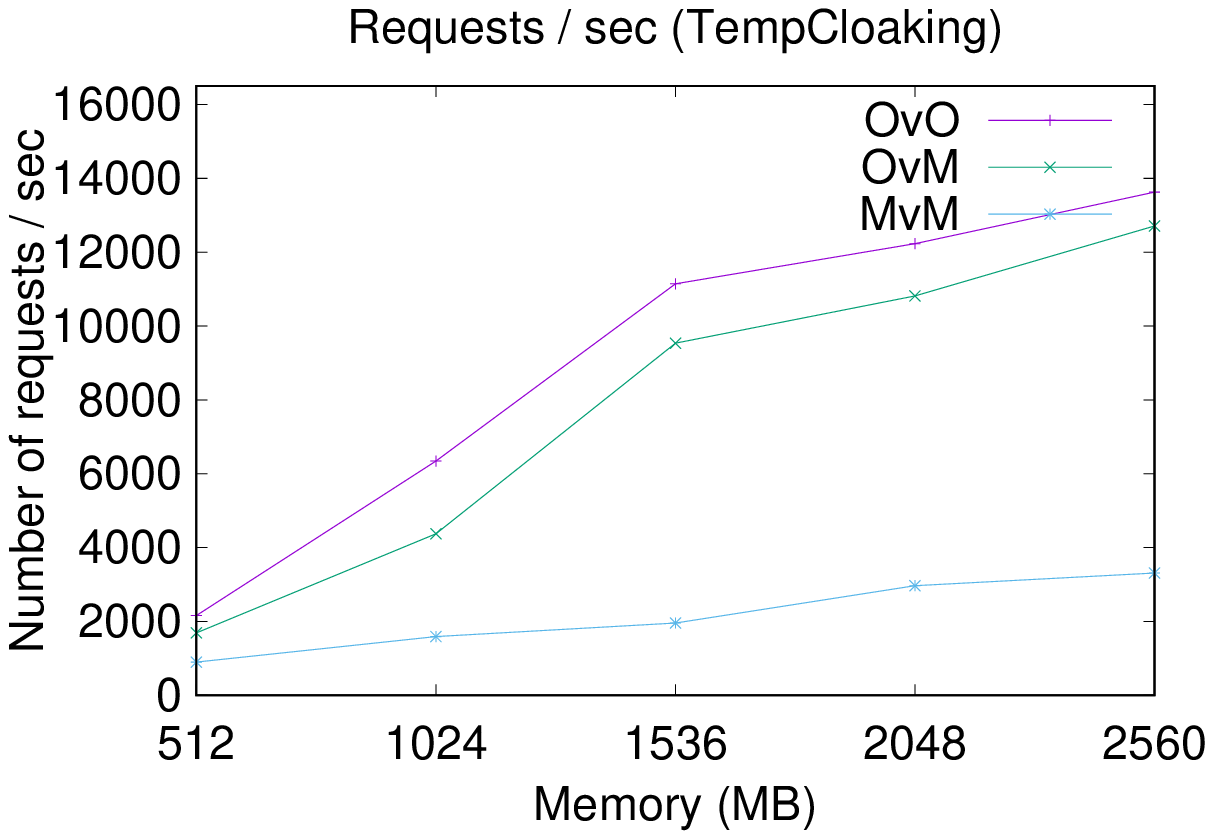}\label{fig:tempcloakreqsec}}
\hfill
\subfigure[Cloaking]{
\includegraphics[width=0.215\linewidth]{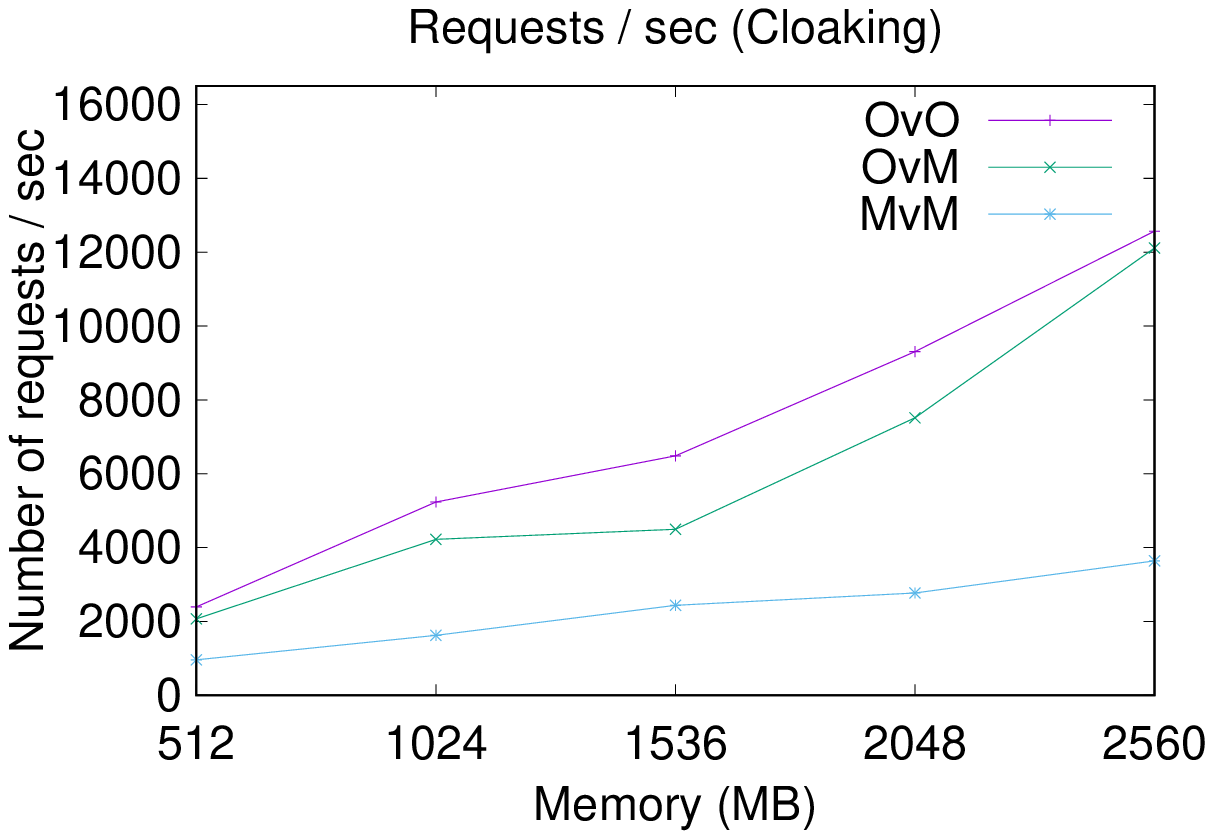}\label{fig:spatcloakreqsec}}
\hfill
\caption{Requests / sec per privacy operation}
\end{minipage}
\end{figure*}

{\bf Data Description. }
To validate our analysis we utilized a real-world dataset from the Foursquare location-based service. The Foursquare dataset includes check-in data in New York city collected from Foursquare from 12 April 2012 to 16 February 2013 \cite{yang2014modeling}. The dataset contains 227428 check-ins and 1083 users, where the users were anonymized for privacy reasons. We selected 60\% of the dataset's users to train the appropriate MPs and the remaining 40\% users as the test set. 
This represents a range of cases in a real system where 
a data analyst can have a range of knowledge about users and their respective trajectories, but it is not possible at all times to have a global view of the users.

\begin{figure*}[htp]
\begin{minipage}{\linewidth}
\centering
\subfigure[Path Confusion]{
\includegraphics[width=0.215\linewidth]{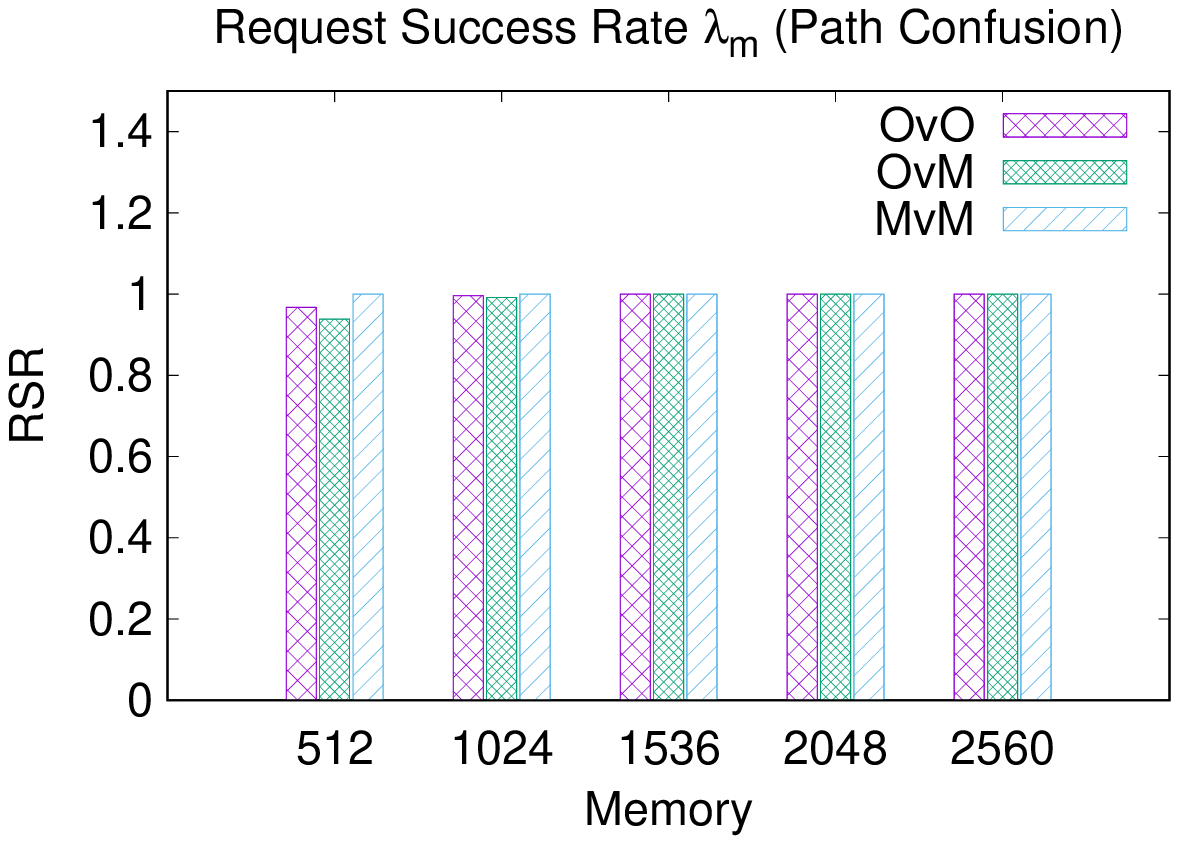}\label{fig:pathconfsucc}}
\hfill
\subfigure[Dummy Locations]{
\includegraphics[width=0.215\linewidth]{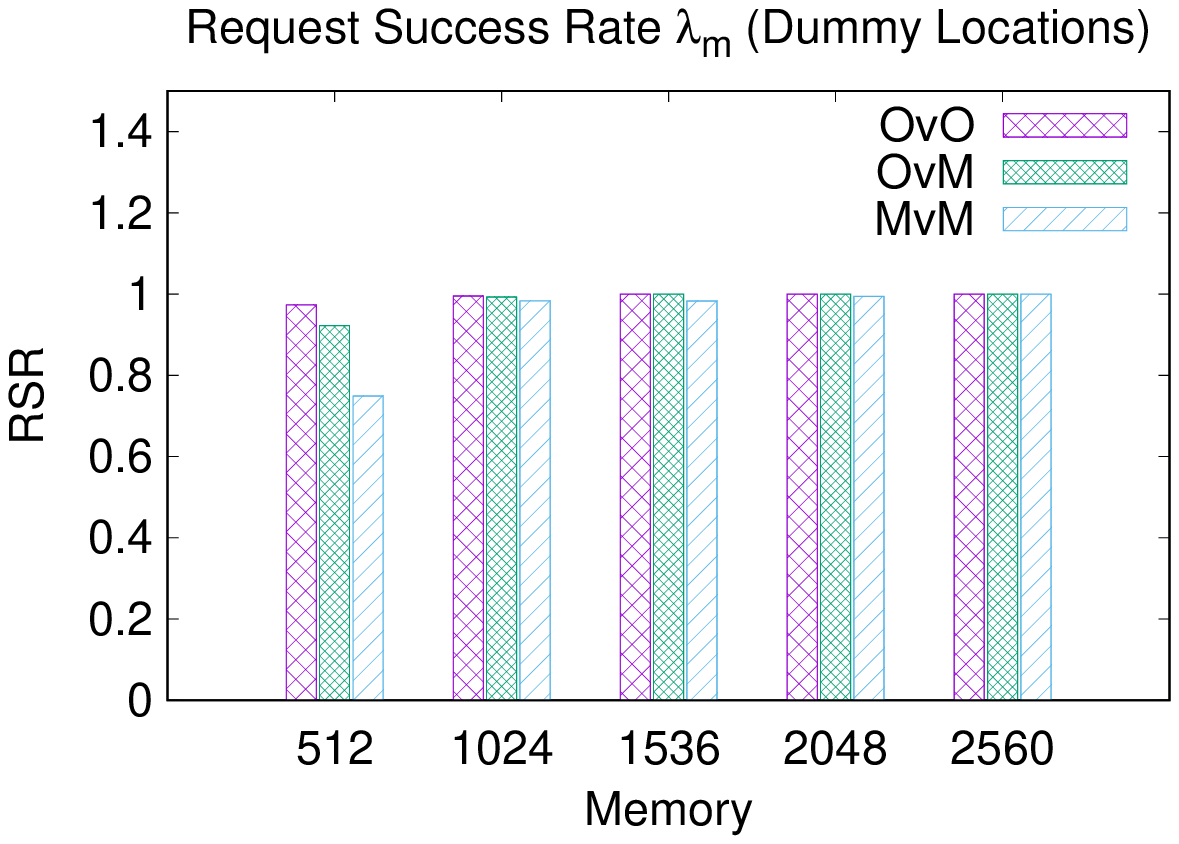}\label{fig:dummysucc}}
\hfill
\subfigure[TempCloaking]{
\includegraphics[width=0.215\linewidth]{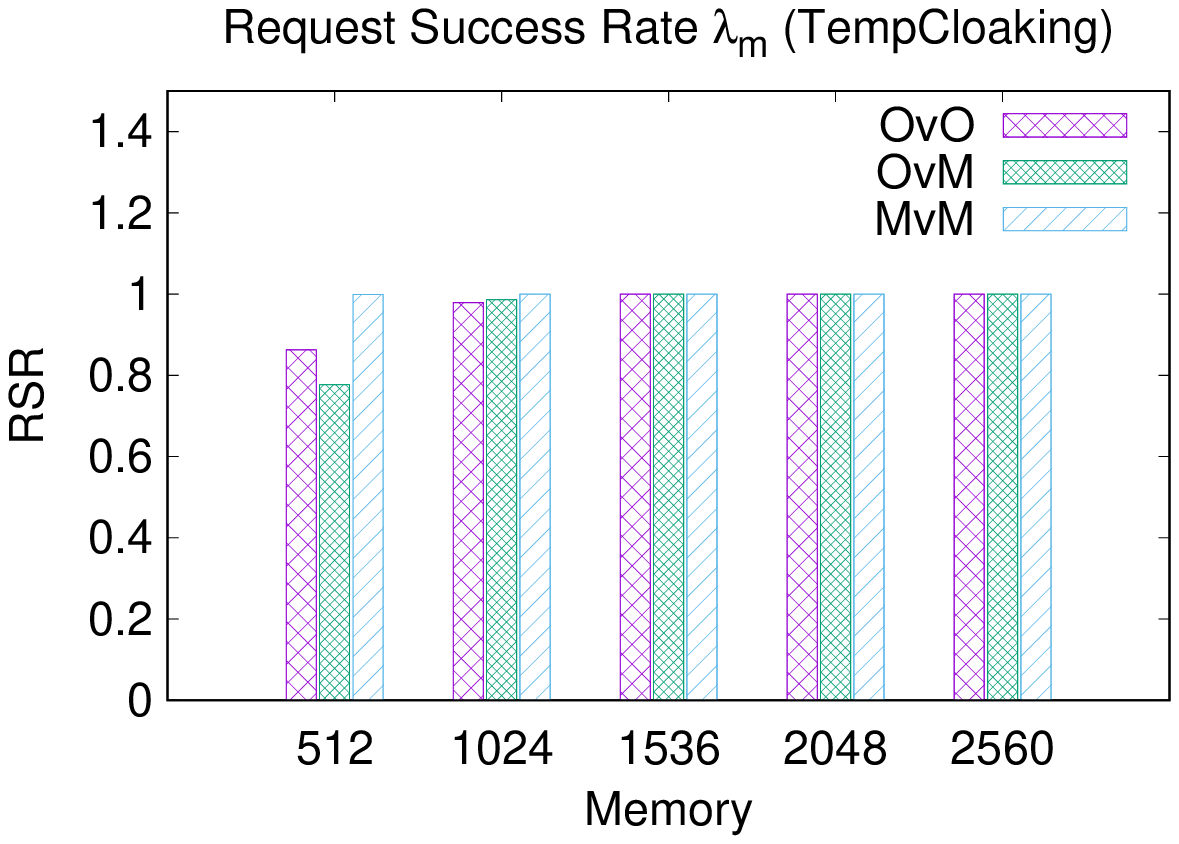}\label{fig:tempcloaksucc}}
\hfill
\subfigure[Cloaking]{
\includegraphics[width=0.215\linewidth]{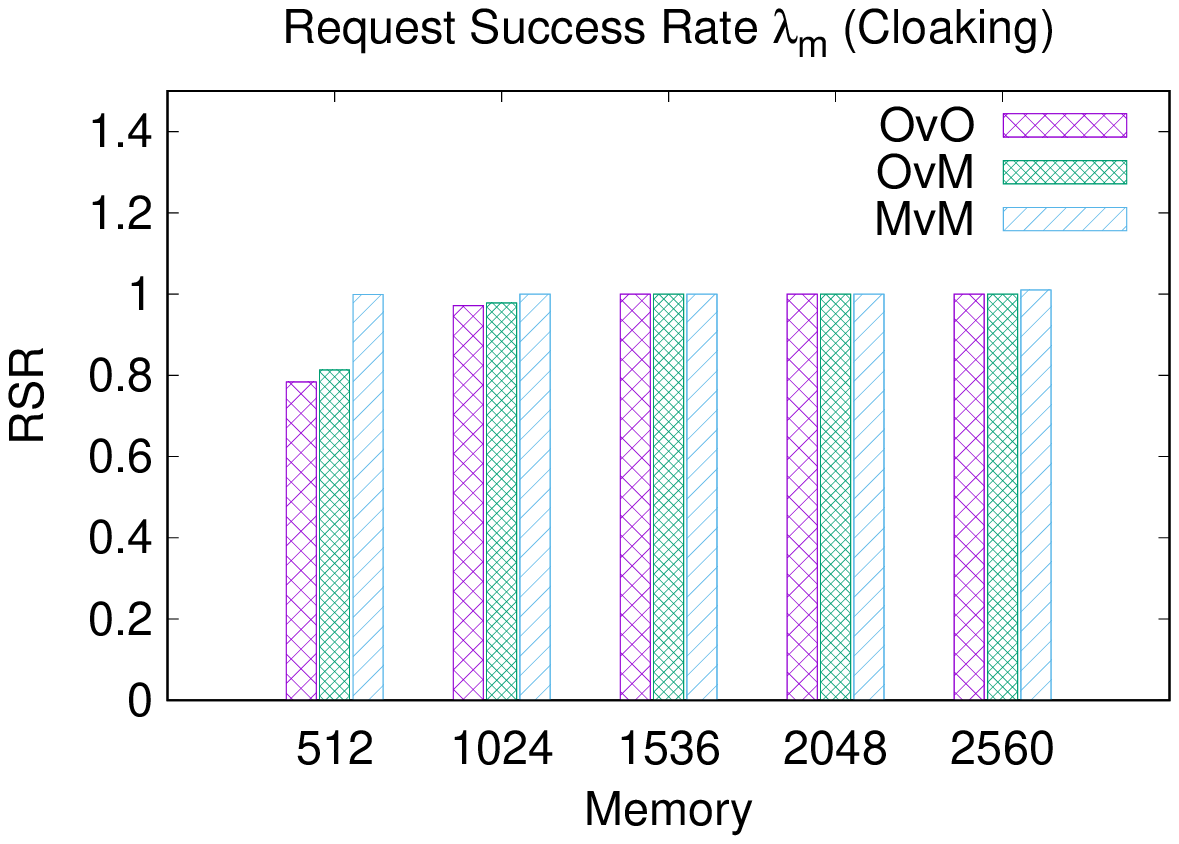}\label{fig:spatcloaksucc}}
\hfill
\caption{Requests Success Rate ($\lambda_{m}$) per privacy operation}
\end{minipage}
\end{figure*}

{\bf Setup. }
We deployed OpenFaaS in our cluster, which consists of 5 nodes (Intel(R) Core(TM) i7 3770 CPU @3.40GHz, 16GB RAM, Ubuntu 16.04 LTS). We used Docker Swarm as the orchestrator and we deactivated Prometheus AlertManager in order to deactivate autoscaling of functions. We used $Hey$ (https://github.com/rakyll/hey) for traffic measurements. We tested three different workload scenarios with five different memory allocation setups: (512, 1024, 1536, 2048 and 2560 (in MB)). In OpenFaaS, we setup our images to allocate 512MB of memory and we used replication in order to increase the total allocated memory. That is, each machine will host a container of each function with 512MB of allocated memory (so 1024MB means 2 replicas in 2 machines, 1536MB 3 replicas in 3 machines etc.).

{\bf Mobility Profile Coverage Rate.}
We introduce a novel metric, namely mobility profile coverage rate, for capturing the percentage of users that are associated with compiled mobility profiles. The metric considers how similar the trajectory of a user is compared to the compiled mobility profiles. Formally,
$    
CR(score(\forall tr_{u}^{l},\mathcal{B}) \geq \delta)=\frac{\#users(score(tr_{u}^{l},\mathcal{B}) \geq \delta)}{TotalUsers}
$
where $\forall tr_{u}^{l}$ denotes every user that has $score(tr_{u}^{l},\mathcal{B})$ above the threshold $\delta$ and \# denotes the number of users.

{\bf Analysis. }
% \begin{figure}[t!]
% \centering
% \begin{minipage}{0.49\linewidth}
% \centering
% \includegraphics[width=\linewidth]{results/coverageRate.eps}
% \caption{Cov. Rate vs Traj. Similarity}\label{fig:covRate}
% \end{minipage}
% \begin{minipage}{0.49\linewidth}
% \centering
% \includegraphics[width=\linewidth]{results/coverageRateAbs.eps}
% \caption{Capt. Users By MPs.}\label{fig:covRateAbs}
% \end{minipage}
% \end{figure}
%Figure \ref{fig:catpop} illustrates the most popular categories for user activities. It is clear that most people share reports while commuting, visiting the gym and while they are entertained. 
Figures \ref{fig:considertraj} \& \ref{fig:avgchain} illustrate the number of trajectories considered and what is the average length of the training trajectories. As we may conclude, selecting an $8-hour$ time window length is reasonable for selecting the trajectories, since it provides a good balance between the length of trajectories to explore and the number of trajectories.
%Vana3: giati 8 time window einai reasonable?
%Dimitris-UBI: Addressed.
In Figure \ref{fig:covRate} we draw the coverage rate for different degrees of trajectory similarity. As we may observe, as we set higher levels of similarity with the mobility profiles, the coverage rate decreases. However, even for only 30\% of trajectory similarity over 60\% of users are captured by the mobility profiles. In Figure \ref{fig:covRateAbs} we draw the number of users captured by the mobility profiles. We observe that even for the case of 50\% similarity, the number of captured users is 200 out of the 1084 users, which consists a considerable number of users. Thus, we conclude that mobility profiles may expose the privacy of a large number of users even with low mobility pattern similarity.

{\bf Baselines. }
We evaluated the performance of $TP^{3}$'s privacy operations compared to a state-of-the-art technique, namely SmartMask \cite{li2016privacy}, which applies a location obfuscation technique that assigns a spatio-temporal data report(a check-in), to the nearest, in terms of distance, point-of-interest.

\subsection{Experimental Workloads}
To show the benefits of employing the serverless model for the privacy functions and different user inputs, we used Hey to generate 20000 HTTP requests using as a payload either single trajectories or mobility profiles, without any rate limitation to stress $TP^3$ to its limits and test each privacy function independently. We setup three real-world scenarios, in which, each user of the test set, is willing to share either a single trajectory or the whole set of the trajectories that consist his mobility profile. The experimental results draw the average value for each one of the examined metrics from all the users in the test set.
%We measured the total time the system took to process the requests, from which we derived the average response time per request. Also we calculated the total requests per second that our system successfully processed. 

%Vana3: ta parakatw scennario thelounn  douleia.. xreiazetai mia protash se kathe scenario ti theloume na deixoume sto anntistoixo scenario.. epishs den tsekaroume mono 1 mobility profile alla ola ta mobile profiles, swsta? mporeite na to deite na ta grapsoume ligo pio kala?
%Dimitris: Addressed.
{\bf One-vs-One Scenario - (OvO).} In the first scenario, we examine the performance of the system when each user shares a single trajectory in the serverless system, evaluated against the most frequent mobility profile encountered in our system. 

{\bf One-vs-Many Scenario - (OvM).} Second, we examine the case in which the user shares a single trajectory which is evaluated against the whole set of mobility profiles compiled by our system. The goal is to capture the possible set of users with whom he may have similar mobility patterns with his trajectory.

\begin{figure*}[htp]
\begin{minipage}{\linewidth}
\centering
\subfigure[Path Confusion]{
\includegraphics[width=0.215\linewidth]{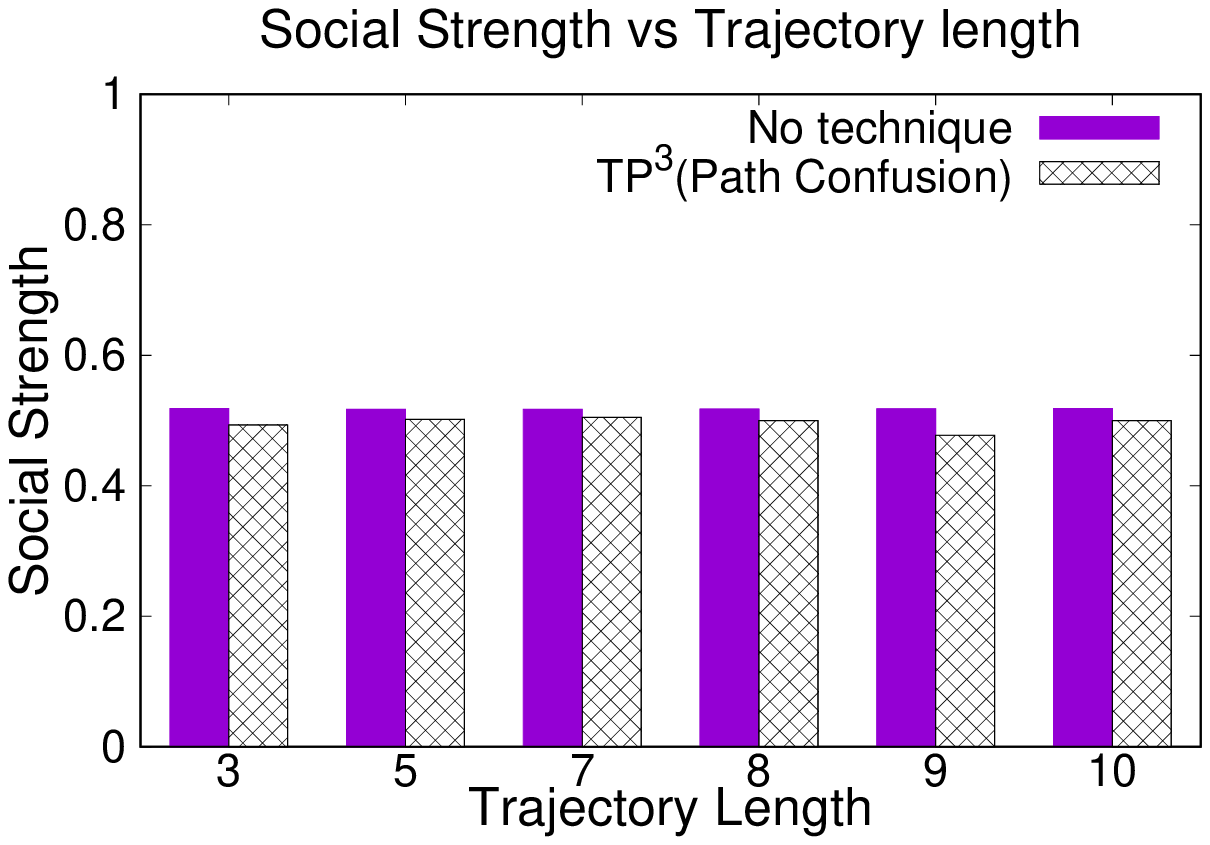}\label{fig:pathconfsimkappa}}
\hfill
\subfigure[Dummy Locations]{
\includegraphics[width=0.215\linewidth]{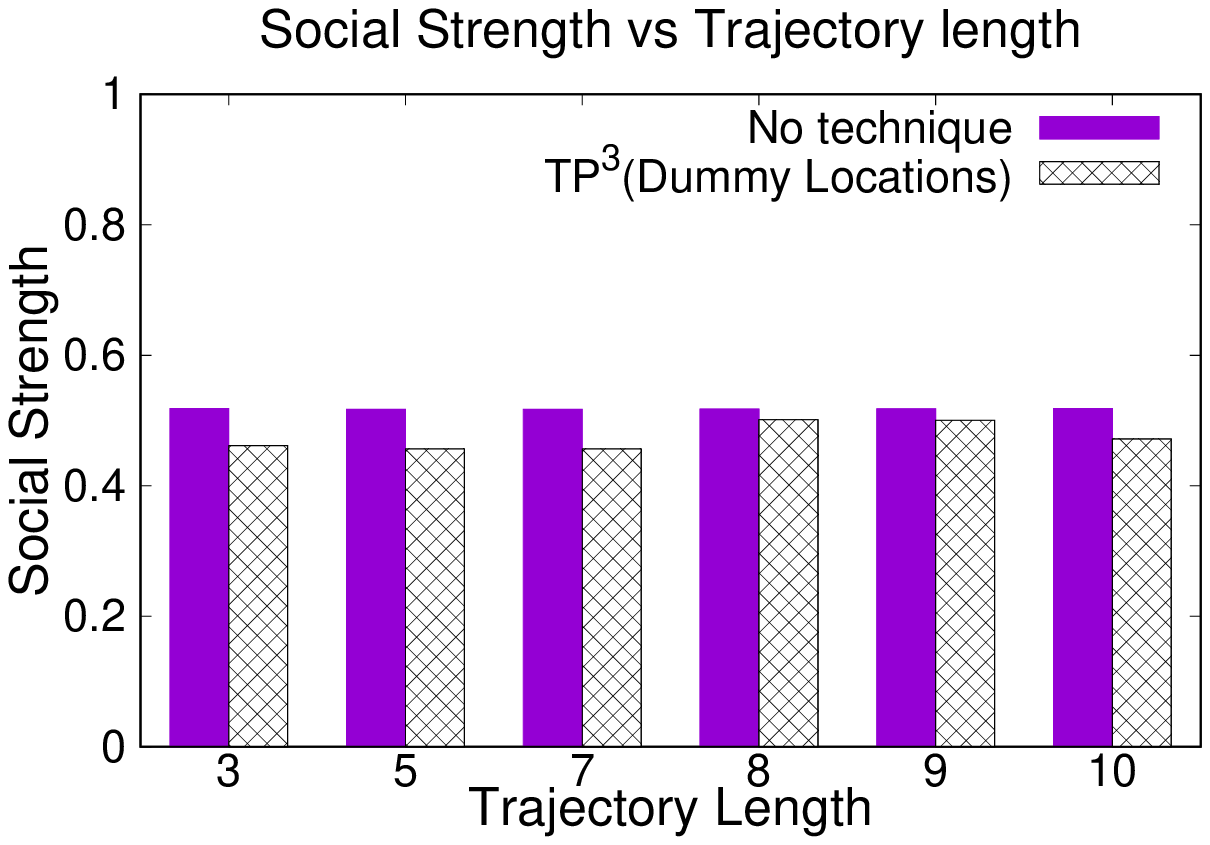}\label{fig:dummysimkappa}}
\hfill
\subfigure[TempCloaking]{
\includegraphics[width=0.215\linewidth]{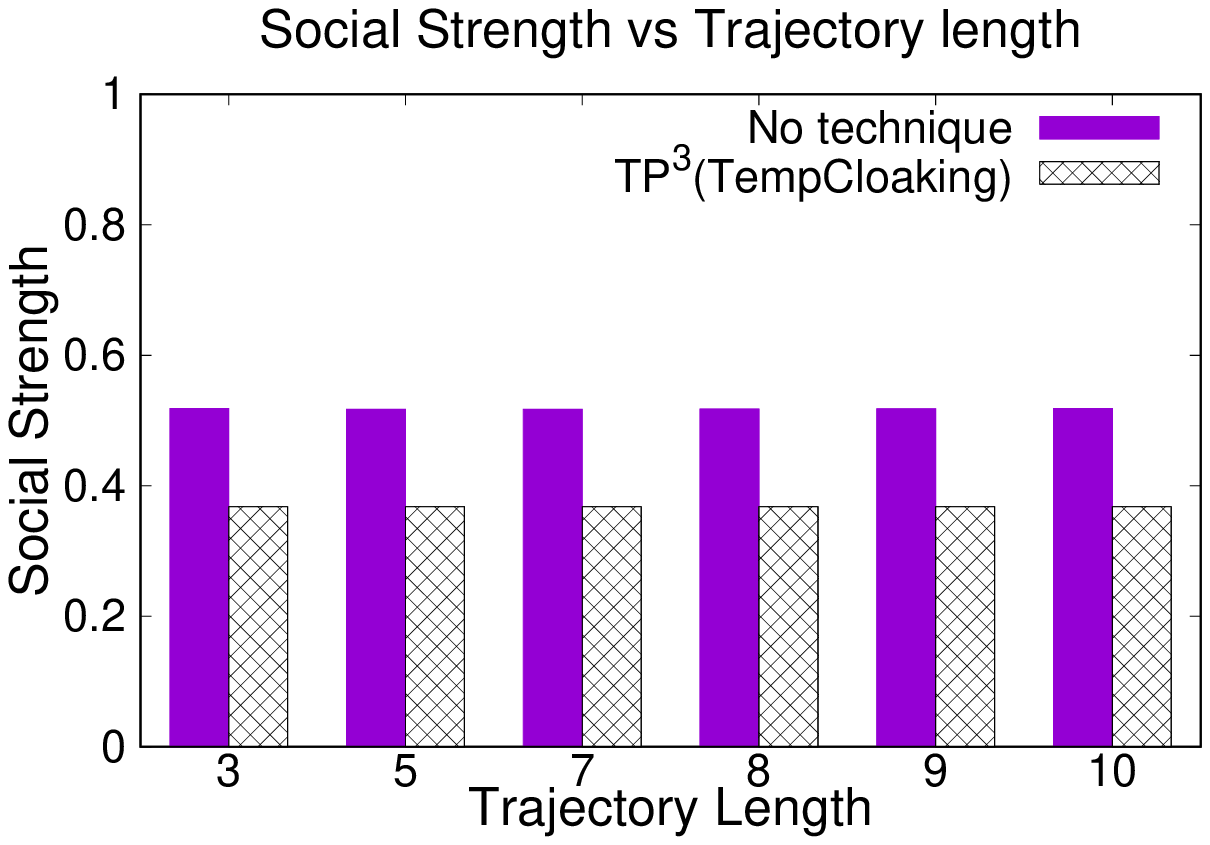}\label{fig:tempcloaksimkappa}}
\hfill
\subfigure[Cloaking]{
\includegraphics[width=0.215\linewidth]{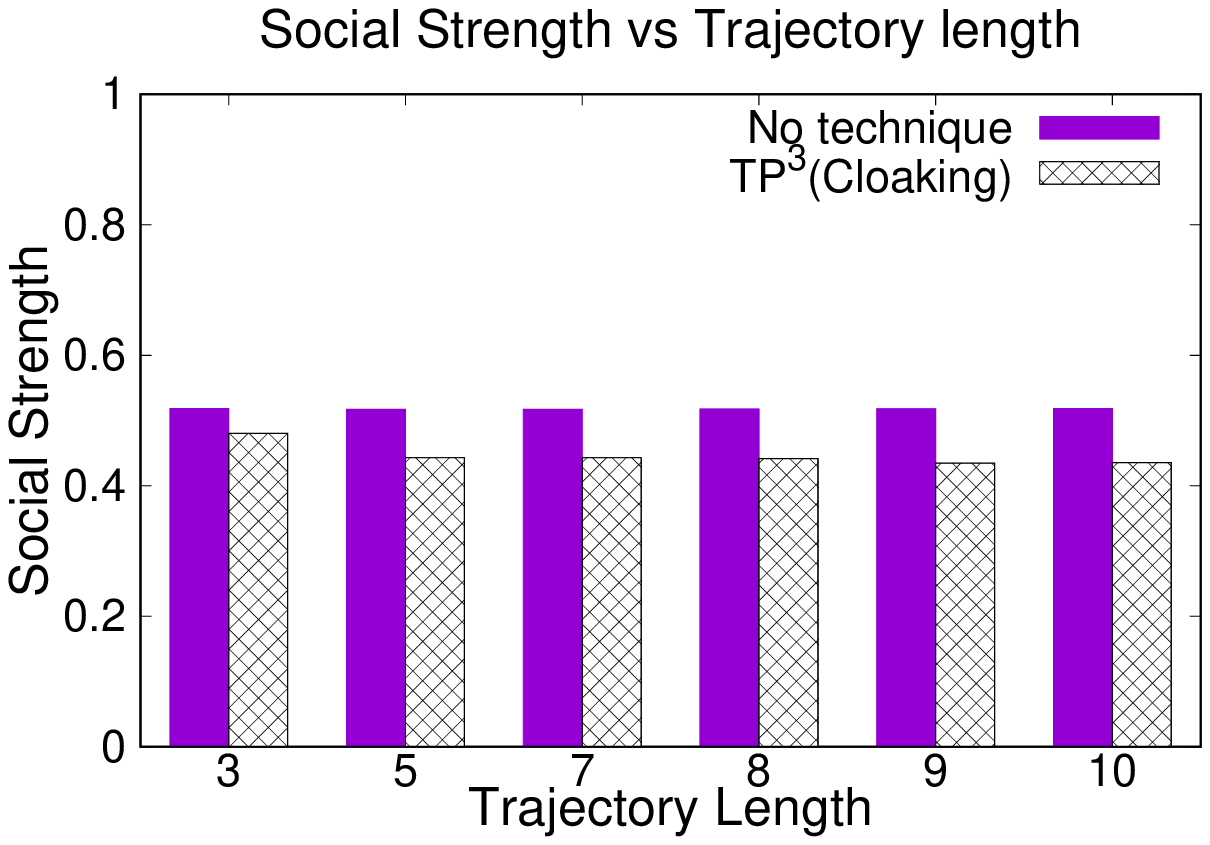}\label{fig:spatcloaksimkappa}}
\hfill
\caption{Social Strength of privacy operations (per length of trajectory)}
\end{minipage}
\end{figure*}

{\bf Many-vs-Many Scenario - (MvM).} Last, we evaluate the case where the user shares his whole set of trajectories, {\it i.e.,} his entire mobility profile, against the whole set of mobility profiles compiled by our $TP^3$ system. Compared to the previous scenario, in this one we aim at capturing all possible users with similar MPs, considering the entire MP of the user.

\subsection{Parameters Examined}

\begin{figure*}[htp]
\begin{minipage}{0.215\linewidth}
\centering
\includegraphics[width=\linewidth]{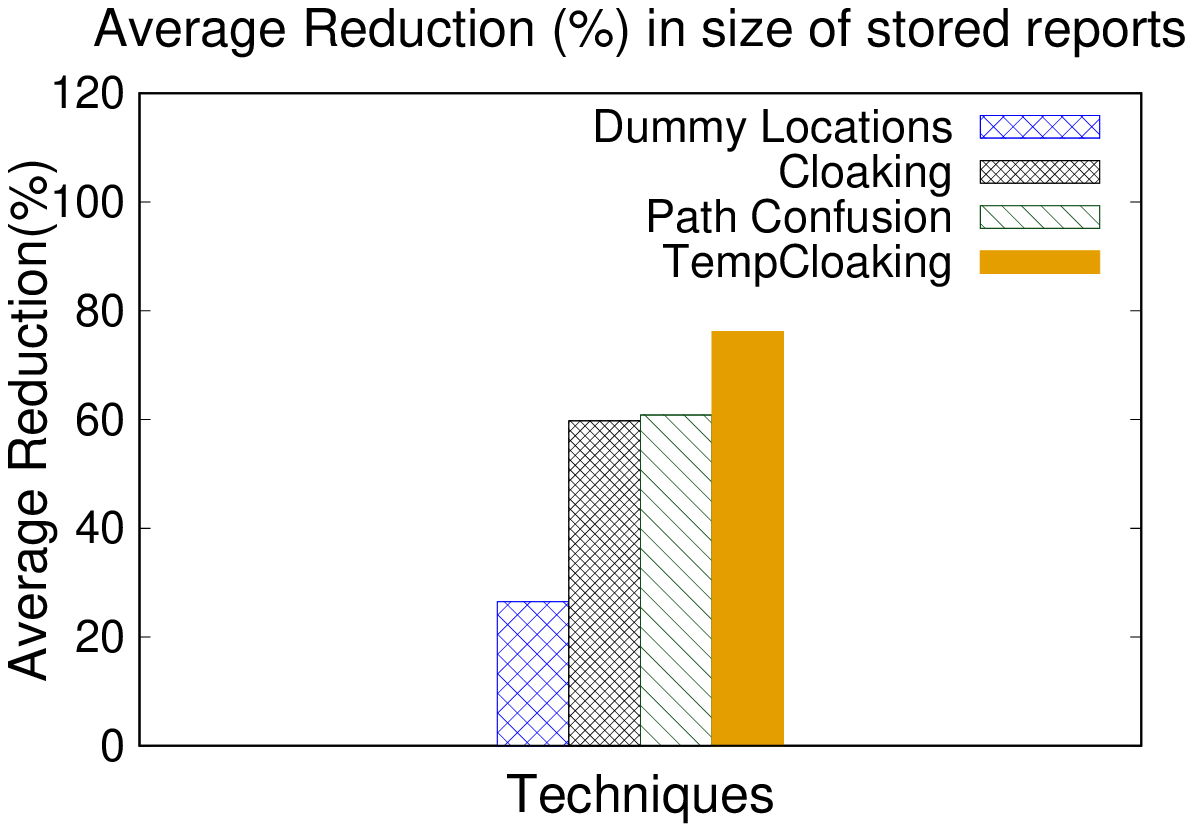}
\caption{Reduction (\%) in size of stored reports}\label{fig:reductionpercent}
\end{minipage}\hfill
\begin{minipage}{0.215\linewidth}
\centering
\includegraphics[width=\linewidth]{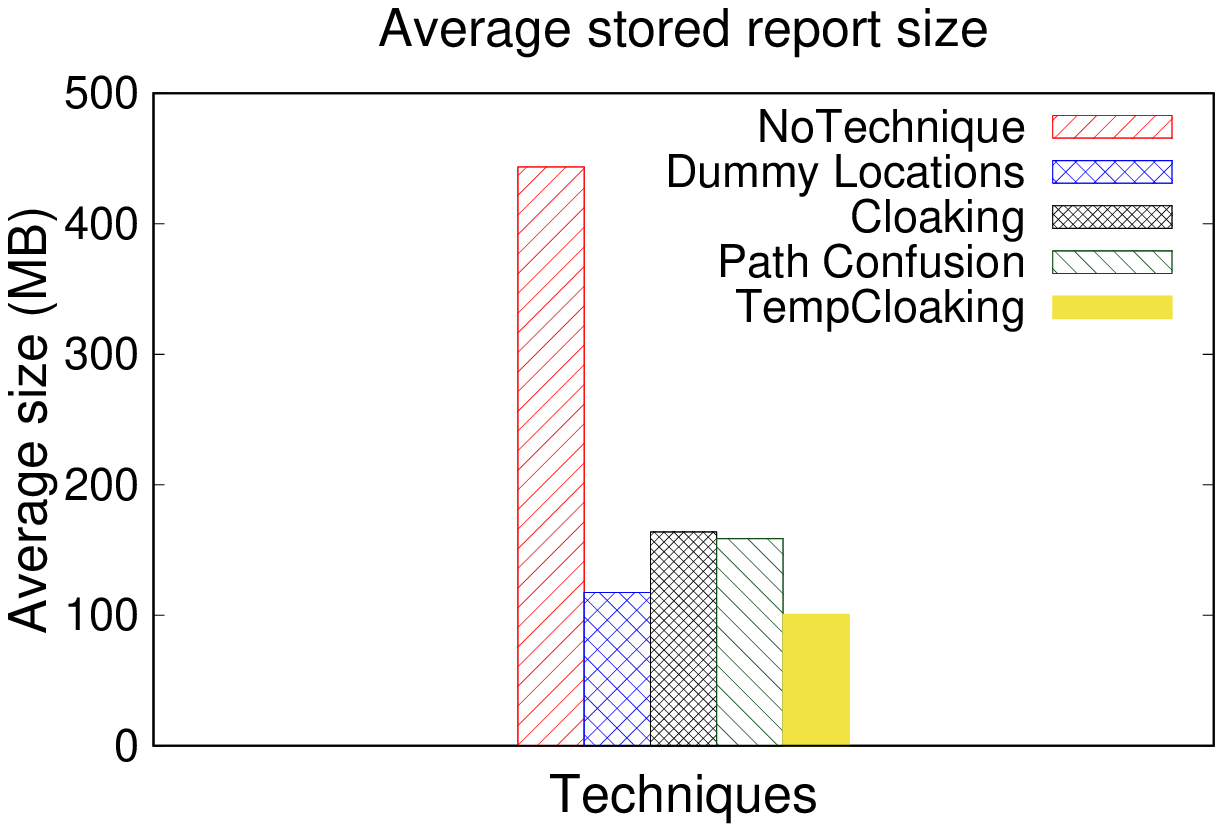}
\caption{Reduction in (MB) of stored reports}\label{fig:reductionabsolut}
\end{minipage}\hfill
\begin{minipage}{0.215\linewidth}
\centering
\includegraphics[width=\linewidth]{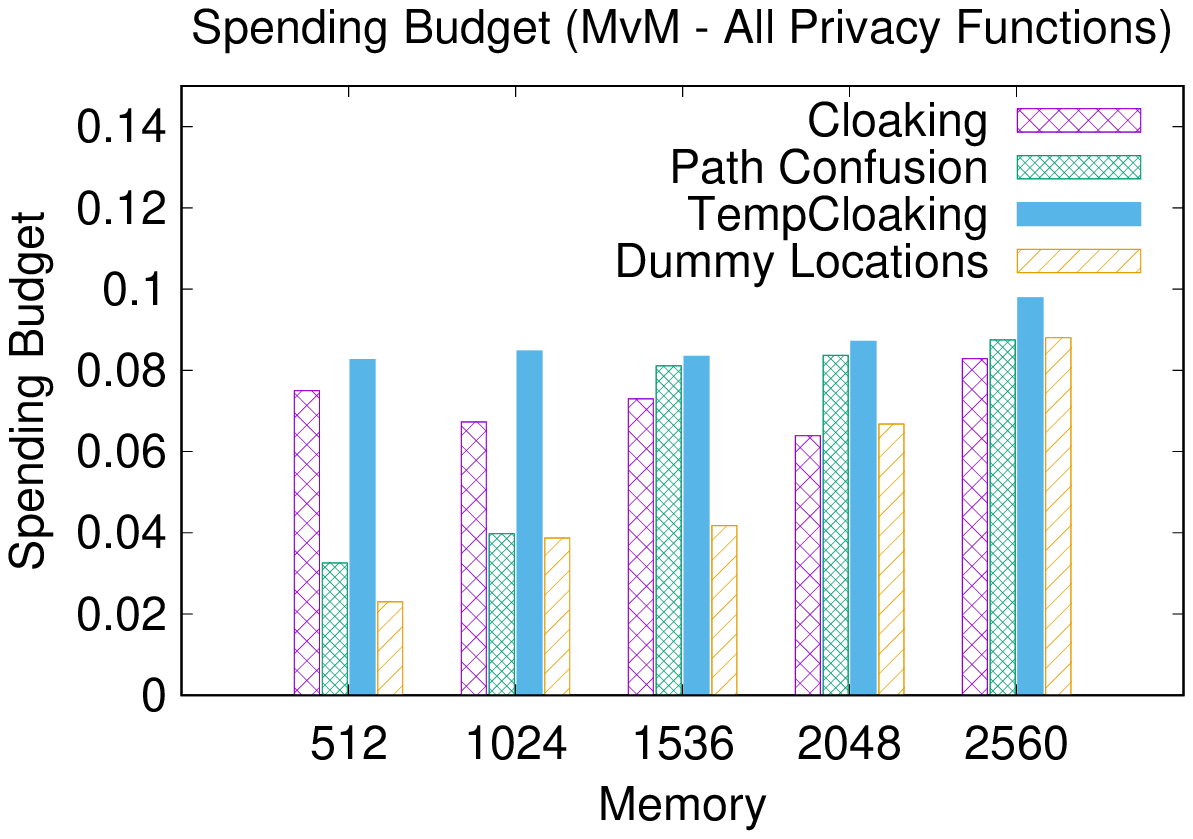}
\caption{Spending Budget for all privacy functions in MvM scenario}\label{fig:allcost}
\end{minipage}\hfill
\begin{minipage}{0.215\linewidth}
\centering
\includegraphics[width=\linewidth]{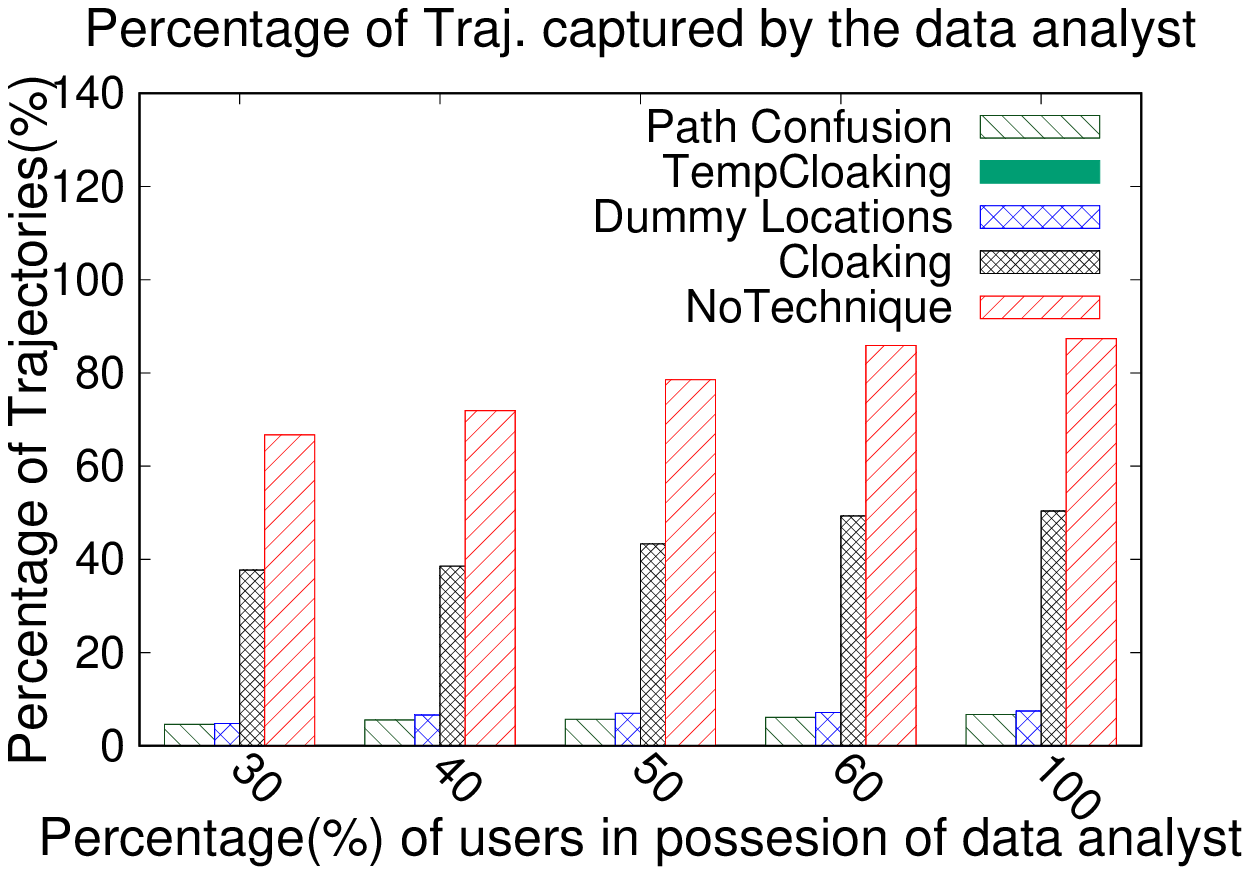}
\caption{Similarity vs Users in possession of the data analyst app}\label{fig:adverpos}
\end{minipage}\hfill
\end{figure*}

{\bf 1)Average Response Time vs Memory.}
In Figures \ref{fig:pathconfavgrec}, \ref{fig:dummyavgrec}, \ref{fig:tempcloakavgrec} and \ref{fig:spatcloakavgrec} we illustrate the Average Response Time of our system which denotes the flexibility of the serverless model with respect to the memory allocation. Our results depict the impact of having flexible memory allocations in the amount of time required to respond to a request. An interesting finding that we observe, is that, increasing the total amount of allocated memory, this results to a \emph{significant} decrease of the average response time for all the privacy functions in all scenarios.

% \begin{figure*}[htp]
% \begin{minipage}{\linewidth}
% \centering
% \subfigure[Path Confusion]{
% \includegraphics[width=0.18\linewidth]{results/theta/pathThetaLines}\label{fig:pathconfsimtheta}}
% \hfill
% \subfigure[Dummy Locations]{
% \includegraphics[width=0.18\linewidth]{results/theta/dummyThetaLines}\label{fig:dummysimtheta}}
% \hfill
% \subfigure[TempCloaking]{
% \includegraphics[width=0.18\linewidth]{results/theta/tempThetaLines}\label{fig:tempcloaksimtheta}}
% \hfill
% \subfigure[Cloaking]{
% \includegraphics[width=0.18\linewidth]{results/theta/cloakThetaLines}\label{fig:spatcloaksimtheta}}
% \hfill
% \caption{Average similarity of privacy operations (per $\theta$-value)}
% \end{minipage}
% \end{figure*}

\begin{figure}[htp]
\begin{minipage}{0.45\linewidth}
\centering
\includegraphics[width=\linewidth]{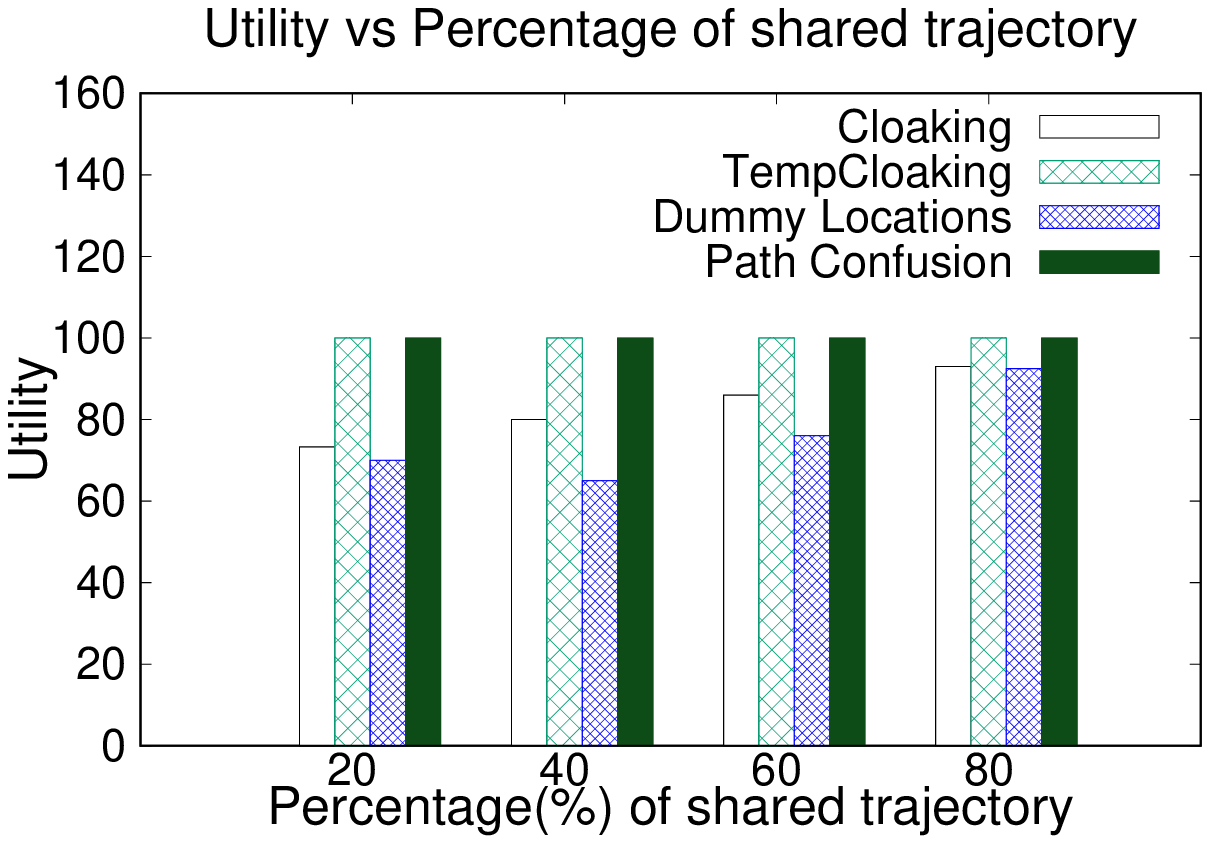}
\caption{Utility vs Percentage of shared trajectory}\label{fig:util}
\end{minipage}\hfill
\begin{minipage}{0.45\linewidth}
\centering
\includegraphics[width=\linewidth]{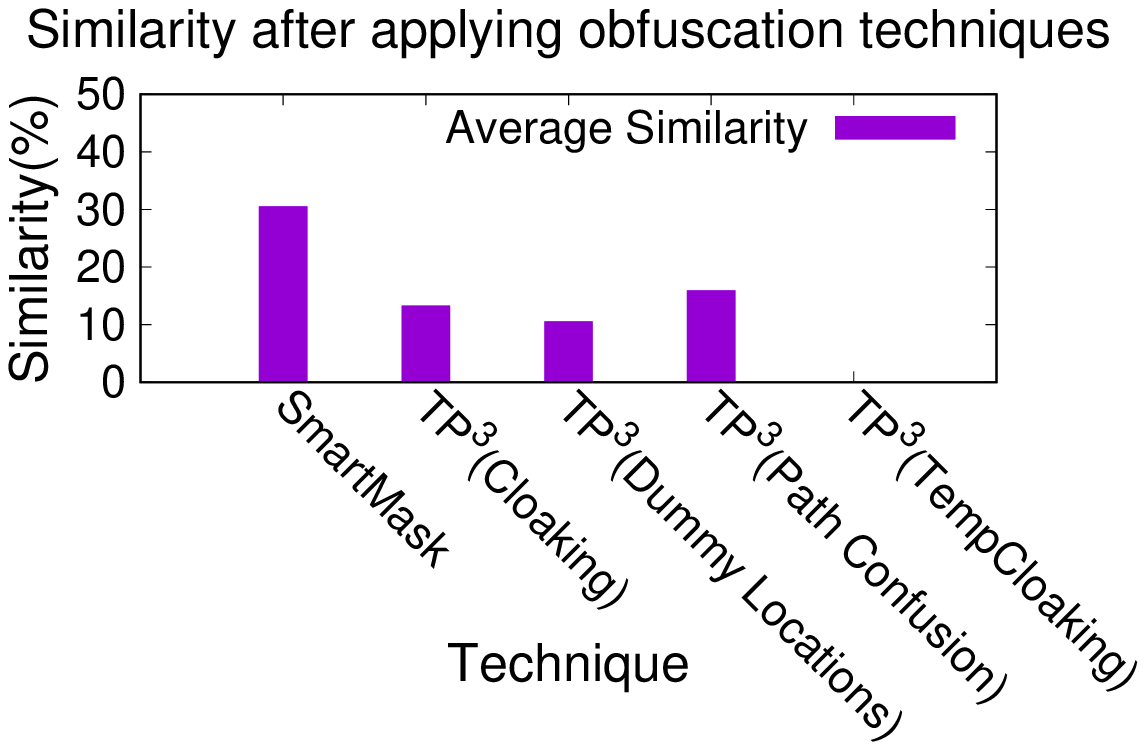}
\caption{Comparison with state-of-the-art}\label{fig:comparisons}
\end{minipage}
\end{figure}

{\bf 2) Requests/sec vs Memory.}
% We measured the performance benefit of the serverless model in terms of requests per second over different memory allocations. 
Figures \ref{fig:pathconreqsec}, \ref{fig:dummyreqsec}, \ref{fig:tempcloakreqsec} and \ref{fig:spatcloakreqsec} depict our initial intuition that, by having flexible memory allocations using a serverless model, we should expect high increase in the number of requests served per sec. We reason these results to the fact that with higher memory the system can handle more concurrent HTTP requests and we can safely conclude that it is beneficial to employ a serverless model for serving multiple concurrent requests.

{\bf 3) Requests Success Rate vs Memory.} 
In Figures \ref{fig:pathconfsucc}, \ref{fig:dummysucc}, \ref{fig:tempcloaksucc} and \ref{fig:spatcloaksucc}, we draw the ratio of successful requests served using a serverless model towards the total number of requests received for a specific memory allocation. The results show that as we increase the allocated memory, and consequently the number of replicas, the ratio increases to a value equal to one, meaning all user requests are successfully served by $TP^3$. This result is expected due to our setup, since the Docker Swarm load balances the traffic across all active replicas. 
% We observed some dropped requests for a memory allocation of 512MB for memory intensive techniques (Dummy Locations and TempCloaking), which required to be served by a single replica. 
Overall, the results show that it is beneficial to employ a serverless model for maximizing the system's performance.

{\bf 4) Trajectory length vs Social Strength. } We evaluated the performance of each privacy operation for reducing the social strength considering trajectories of different length. We considered trajectories with length equal or greater than a varying trajectories length parameter. Figures \ref{fig:pathconfsimkappa}, \ref{fig:dummysimkappa}, \ref{fig:tempcloaksimkappa} \& \ref{fig:spatcloaksimkappa} illustrate the social strength in logarithmic scale observed after applying each privacy operation for different values of trajectory length. An interesting finding is that the Dummy Locations technique outperforms the Cloaking technique for a trajectory length equal to 3. This is due to the fact that the Dummy Locations technique better obfuscates the user trajectory, since the random points inserted for this trajectory length, change it significantly. Overall, the TempCloaking and Cloaking privacy operations outperform the other techniques in terms of social strength minimization.

% DIMITRIS: Tha vgaloume auto to experiment kai stin thesi tou tha treksoume gia ta idia theta values, na doume poso meiwnetai to size.

{\bf 5) Coreset size vs Stored Reports.} We investigated the effect of the size of the coresets on the size of stored reports when each privacy operation is applied. From our experimental analysis, we considered a value of $\theta=0.0005$ which is reflected on a moderate level of sampling (and is the default settig in our implementation). Figures \ref{fig:reductionpercent} and  \ref{fig:reductionabsolut} illustrate the performance in terms of stored reports size reduction for each privacy operation. The results depict that the privacy operations allow for reducing the number of stored reports in the users phones significantly. We observe that TempCloaking (76\% size reduction) still outperforms the privacy operations having Path Confusion and Cloaking as runner ups.

{\bf 6) Spending Budget vs Memory.} We examined the spending budget performance against different memory allocations and for every workload scenario. In Figure \ref{fig:allcost}, we draw the spending budget performance for the many-to-many scenario, which is the most heavy in terms of performance, for the different privacy operations and memory allocations. We observe that the spending budget is relatively low compared to having a fixed budget. In addition, we observe that we can vary the spending budget with regards to the serverless privacy function we want to execute, thus providing a tunable-degree of privacy with regard to the execution costs.

{\bf 7) Percentage of data analyst trajectories. } 
% We evaluated the performance of $TP^{3}$'s privacy operations reagarding the percentage of the trajectories the data analyst can have in his possession. 
In Figure \ref{fig:adverpos} we draw the percentage of trajectories captured by the percentage of users for whom the data analyst has compiled MPs, with no privacy technique applied and after using $TP^{3}$'s privacy operations. We observe that as the percentage of users in possession of the data analyst  increases, the percentage of trajectories captured also increases for all the privacy operations but still remains under 50\%. That is, since the data analyst has the 100\% of the users, only $\sim$45\% of user trajectories can be detected if only Cloaking technique is selected from all users to be applied. We also observe that TempCloaking technique totally minimizes the similarity of user trajectories with a MP due to the fact that it totally changes the nature of the user's trajectories. Path Confusion and Dummy Locations methods also present good performance and so, we can conclude that $TP^{3}$ can successfully trade-off between different levels of privacy and desired accuracy of results.

% \begin{figure}[t!]
% \centering
% \begin{minipage}{0.48\linewidth}
% \centering
% \includegraphics[width=\linewidth]{results/utility/utility.eps}
% \caption{Utility vs Percentage of shared trajectory}\label{fig:util}
% \end{minipage}
% \begin{minipage}{0.48\linewidth}
% \centering
% \includegraphics[width=\linewidth]{results/comparison.eps}
% \caption{Comparison with state-of-the-art}\label{fig:comparisons}
% \end{minipage}
% \end{figure}

% \begin{figure}[t!]
% % \begin{minipage}{0.45\linewidth}
% % \centering
% % \includegraphics[width=\linewidth]{results/percentageOfUsers.eps}
% % \caption{Similarity vs Users in possession of adversary}\label{fig:adverpos}
% % \end{minipage}\hfill
% \centering
% \begin{minipage}{0.55\linewidth}
% \centering
% \includegraphics[width=\linewidth]{results/comparison.eps}
% \caption{Comparison with state-of-the-art}\label{fig:comparisons}
% \end{minipage}
% \end{figure}

{\bf 8) Utility vs Percentage of Shared Trajectory. } We investigated the balance between data quality and privacy, when each one of $TP^{3}$'s privacy operations is applied. 
Once a privacy operation has been selected, the goal is to apply it while preserving as much data utility as possible.
In Figure \ref{fig:util} we draw the utility for the different privacy operations applied. The x-axis denotes the percentage of the trajectory that has already been published without applying any privacy operation. The published data are considered useful when the utility equals 100\%. Path Confusion and Temporal Cloaking have utility 100\%, since in the Foursquare location-based application they affect only the times of the reports rather than the corresponding locations.
%and thus time and direction, do not play a crucial role. 
Overall, we safely conclude that $TP^{3}$ succeeds in maintaining a balance between the data accuracy and privacy.

{\bf 9) Comparison with state-of-the-art. } In Figure \ref{fig:comparisons} we draw the percentage of similarity for the different privacy operations applied. We observe that $TP^{3}$ outperforms SmartMask, since for every provided privacy operation by $TP^{3}$, it results in lower similarity (it performs 47\% better than SmartMask). This implies that $TP^{3}$ is practical and efficient for protecting against social link exploitation attacks.

%% file: related.tex
\section{Related Work}

{\bf Privacy Models.} Privacy preservation is not a new area and approaches have been proposed in the literature\cite{backes2017walk2friends,pham2013ebm,AhujaGS19,yao2019publishing,kalnis2007preventing}. However, these works have several limitations.
In \cite{backes2017walk2friends}, they propose an attack that predicts social links between users, but it does not consider trajectories nor focuses on how a user associates to a group of users based on his mobility patterns. The authors of \cite{pham2013ebm} aim at understanding the significance of a location visited by a user, which is encapsulated in the number of visits to a specific location. However, it is limited since the focus is on single locations rather than trajectories. In \cite{yao2019publishing}, the authors focus on securing the sensitive attributes of each location visited by applying l-diversity whereas in our approach we provide different privacy operations tailored to user needs. The authors of \cite{AhujaGS19} focus on Geo-Indistinguishability for single locations than trajectories as we do in our work.
Finally, in \cite{kalnis2007preventing}, the authors proposed transformations based on the K-anonymity concept for user locations, without considering the users' mobility patterns nor possible social ties, which is the focus of our work. %Finally, the authors of \cite{xiao2015protecting} propose a systematic solution to preserve location privacy, without taking into consideration how user trajectories can be affected by the social connections as we do in our work. 

%% file: conclusion.tex
\section{Conclusions}

In this paper we presented $TP^{3}$, a privacy preservation system for trajectory analytics. We have modeled a new type of attack considering how social ties shape human mobility. Our proposed system employs the serverless paradigm and manages to balance the trade-off between maximizing the overall performance and minimizing the operational costs, while requiring low maintenance and administration from the cloud provider. $TP^{3}$ runs in concert with state-of-the-art trajectory analytics apps. Our experimental evaluation, compared to state of the art schemes, illustrates a reduction of at least 47\% in user privacy exposure, providing a tunable degree of privacy preservation with high system performance for users and low costs for cloud providers.